\documentclass[prd,superscriptaddress,nofootinbib,floatfix,twocolumn]{revtex4}

\usepackage{bm,amsmath,amssymb,natbib}
\usepackage{ifthen}
\usepackage{aas_macros}
\usepackage{graphicx}
\usepackage[svgnames]{xcolor}

\newboolean{showcomments}
\setboolean{showcomments}{true}
\newcommand{\antcomment}[1]{\ifthenelse{\boolean{showcomments}}{\footnote{\color{DarkGreen} #1}}{}}
\newcommand{\riccomment}[1]{\ifthenelse{\boolean{showcomments}}{{\footnote{Maroon} #1}}{}}

\newcommand{\changeR}[1]{#1}

\newcommand{\clh}{\mathcal{H}}

\newcommand{\Mpc}{\text{Mpc}}

\newcommand\pdiff[2]{\frac{\partial #1}{\partial #2}}

\newcommand\del{\nabla}

\newcommand{\vv}{\mathbf{v}}

\newcommand{\vx}{\mathbf{x}}

\newcommand{\vk}{\mathbf{k}}

\newcommand{\khat}{\hat{k}}

\newcommand{\brsc}[1]{{\ensuremath{{\scriptscriptstyle \left(#1\right)}}}}

\DeclareMathOperator{\sinc}{sinc}

\newcommand{\msolar}{M_\odot}

\usepackage{hyperref}

\begin{document}

\title{Constraining Primordial Magnetism}

\author{J. Richard Shaw}
\email{jrs65@cita.utoronto.ca}

\affiliation{Kavli Institute for Cosmology, Madingley Road, Cambridge, CB3 0HA, UK.}
\affiliation{Canadian Institute for Theoretical Astrophysics, 60 St George
  Street, Toronto, M5S 3H8, Canada}

\author{Antony Lewis}
\homepage{http://cosmologist.info}
\affiliation{Astronomy Centre, University of Sussex, Brighton, BN1 9QH, UK}

\date{\today}
\begin{abstract}
  Primordial magnetic fields could provide an explanation for the
  galactic magnetic fields observed today, in which case they may also
  leave interesting signals in the CMB and the small-scale matter
  power spectrum. We discuss how to approximately calculate the
  important non-linear magnetic effects within the guise of linear
  perturbation theory, and calculate the matter and CMB power spectra
  including the SZ contribution. We then use various cosmological
  datasets to constrain the form of the magnetic field power
  spectrum. Using solely large-scale CMB data (WMAP7, QUaD and ACBAR)
  we find a 95\% CL on the variance of the magnetic field at $1
  \:\Mpc$ of $B_\lambda < 6.4 \: \mathrm{nG}$. When we include SPT
  data to constrain the SZ effect, we find a revised limit of
  $B_\lambda < 4.1 \: \mathrm{nG}$. The addition of SDSS
  Lyman-$\alpha$ data lowers this limit even further, roughly
  constraining the magnetic field to $B_\lambda < 1.3 \: \mathrm{nG}$.
\end{abstract}


\maketitle
\section{Introduction}

Magnetic fields are observed at the $10^{-6}$G level in galaxies and clusters,
not just in the local Universe, but up to a redshift of $z \sim 0.7$--$2$
\cite{Widrow:2002}. Production of magnetic fields within a formed galaxy is
extremely difficult, and the consensus is that they are amplified from pre-
galactic seed fields \cite{Kulsrud:2008}. One interesting possibility is that
these seed fields were primordial in origin, formed in the very early
Universe. \changeR{These seed field are thought to have originated either in a
phase transition in the early universe, or produced directly during inflation.
However, these scenarios have problems, the former struggles to produce any
significant magnetic fields with the correlation lengths observed
\cite{Hogan:1983,Durrer:2003}, whilst many models of inflationary
magnetogenesis must produce fields of limited amplitude in order that back-
reaction does not halt inflation \cite{Demozzi:2009}.}

The observation of magnetic fields up to a redshift of $z \sim 2$ are
particularly interesting as the fields have amplitudes ($B \sim 10
\mu\mathrm{G}$) comparable to those observed locally \cite{Bernet:2008}.
This evidence seems to disfavour a large dynamo amplification, requiring a
larger seed field to produce today's magnetic fields. In the extreme case of
there being no dynamo amplification, adiabatic contraction alone could amplify a
pre-galactic field of around $1\: \mathrm{nG}$ (comoving) to the required level.

A stochastic primordial magnetic field not only modifies the standard
evolution of the Universe, but sources additional scalar, vector and tensor
modes, giving rise to both new temperature and polarisation perturbations in
the Cosmic Microwave Background as well as modifying the standard scenario
\cite{Subramanian:1998b,Mack:2002,Kahniashvili:2005,Gopal:2005,faradaypaper,Giovannini:2008,Finelli:2008,Paoletti:2008,Yamazaki:2008,Shaw:2009}, and both these
contributions alter the matter distribution in the local universe \cite{Yamazaki:2006b,Sethi:2008,Yamazaki:2008b,Tashiro:2009,Yamazaki:2010b,Kunze:2011}.
Our aim in this paper is to see what limits current data places on the level
of magnetic fields in the early universe, and whether this is compatible with
a scenario where galactic magnetic fields are seeded by primordial magnetic
fields.

Constraints on the strength of a primordial magnetic field come from many
areas. The expansion rate at Nucleosynthesis places direct limits on the
strength of a magnetic field then of $B \lesssim 1 \:\mathrm{\mu G}$
(comoving) \cite{Matese:1970,Grasso:1996}, though thos can be strengthened by
looking at the constraint on magnetic-sourced gravitational waves
\cite{Caprini:2002} (also see Section \ref{sec:caprini}). \changeR{An indirect
constraint suggests that primordial magnetic fields must be limited to
$B_\lambda \lesssim 10^{-12} \:\mathrm{G}$ to match Faraday rotation measures
in clusters \cite{Dolag:1999,Banerjee:2004}, though this bound is heavily
dependent on the modelling of the magnetohydrodynamics during and after
cluster formation.} Observations of the CMB provide the \changeR{strongest
direct} limits, with previous statistical analyses
\cite{Yamazaki:2006,Yamazaki:2010,Paoletti:2010}, finding field strengths of
several $\mathrm{nG}$ to be consistent with current CMB data, \changeR{with
comparable bounds also coming from limits on the contribution to $\sigma_8$
\cite{Yamazaki:2008b,Yamazaki:2010b}}. These analyses only looked at the CMB
power spectrum and some progress has been made in improving these limits
through higher order moments \cite{Brown:2005} such as the CMB bispectrum
\cite{Seshadri:2009,caprini:2009b}.

This work will depend heavily on the results of a previous paper
\cite{Shaw:2009}, and will use the same conventions and notation. Where it is
necessary to use perturbation theory we use a gauge invariant notation
\cite{Durrer:1994,Durrer:1988} similar to the Conformal Newtonian Gauge. In this
work we limit ourselves to a flat $\Lambda$CDM universe.


\section{Non-linear Field Evolution}
\newcommand{\vB}{\mathbf{B}}
\newcommand{\Bc}{\hat{B}}
\newcommand{\vBc}{\hat{\vB}}

We will consider a stochastic magnetic field $B^i(x^j,\tau)$ generated
by some mechanism in the very early Universe. As for all the periods
of interest the Universe contains a highly ionised plasma, we can use
the MHD equations to describe the behaviour of the magnetic field. At
linear order in the magnetic field
\begin{equation}
\pdiff{}{\tau} \bigl[a^2 B^i(\vx, \tau)\bigr] = 0 \; ,
\end{equation}
and this motivates the definition of a comoving magnetic field,
$\Bc^i(\vx, \tau) = a^2 B^i(\vx, \tau)$ which is time independent at
linear order. Using this we can write the non-zero components of the
magnetic energy-momentum tensor as
\begin{subequations}
\begin{align}
T^0_0 & = - \frac{1}{8\pi a^4} \Bc^2(\vx) \; , \\
T^i_j & = \frac{1}{4\pi a^4} \left(\frac{1}{2} \Bc^2(\vx) \delta^i_j - \Bc^i(\vx) \Bc_j(\vx)\right) \; .
\end{align}
\end{subequations}
As there is no magnetic field on the background, the perturbations of
the stochastic background are manifestly gauge invariant. We construct
density and anisotropic stress perturbations, $\Delta_B$ and $\Pi_B$,
defined by
\begin{subequations}
\begin{align}
T^0_0 & = - \rho_\gamma \Delta_B \; ,\\
T^i_j & = p_\gamma \left(\Delta_B \delta^i_j + \Pi_B{}^i{}_j\right) \; ,
\end{align}
\end{subequations}
where we include the factors of $\rho_\gamma$ and $p_\gamma$ to take
account of the $a^{-4}$ factors. As usual the anisotropic stress $\Pi_B{}^i{}_j$ can be
decomposed into scalar, vector and tensor
contributions.

At higher order the comoving magnetic field obeys
\begin{align}
\label{eq:mhd}
\partial_\tau \Bc^i & = \bigl[\del\times(\vv \times \vBc) + \eta \, \del^2 \vBc\bigr]^i \notag \\
&= \epsilon^{ijk} \epsilon_{klm} \partial_j \bigl( v^l \Bc^m \bigr) + \eta \, \partial_j \partial^j \Bc^i
\end{align}
where $\vv$ is the baryon velocity and $\eta$ is the magnetic diffusivity. In a
highly conductive medium $\eta$ is negligible and so we will set it to zero from
here onwards. As \eqref{eq:mhd} is non-linear the magnetic field evolution
cannot be treated accurately in the standard linearised Einstein-Boltzmann
approach. The standard approach is to separate the higher order evolution of the
magnetic field into two effects that are physically well-motivated, a damping
due to the radiation viscosity, and the effect of the magnetic Jeans length (for
extended discussions, see
\cite{Subramanian:1998,Jedamzik:1998,Kim:1996,Barrow:2007}).  We follow the same
route in this work, modifying the linear evolution equations to capture the
essential physics without resorting to a higher order calculation. Though a more
detailed second order calculation would be desirable, it would be
computationally unfeasible to use this in a Markov-Chain Monte-Carlo analysis of
cosmological data.

The first non-linear contribution we address is the magnetic counterpart of the
Jeans effect. As gravitational collapse causes baryon density perturbations to
grow the pressure in the gas is increased, and eventually halts the growth.
When present, a magnetic field gives an additional contribution to the
pressure. As a baryon overdensity collapses it compresses the large scale field,
generating an increase in the magnetic pressure at the scale of the baryon
perturbation. In a magnetised medium there are multiple modes that could be
excited, but we expect the the Jeans effect to be mediated primarily by the
fastest, the fast magnetosonic mode, which corresponds to the normal acoustic
mode in the limit of small magnetic field \cite{Adams:1996}. The simplest way of
encapsulating this is to consider it as a modification of the baryon sound
speed, to the speed of the fast magnetoacoustic mode \cite{Kim:1996,Adams:1996},
$c_s^2 \rightarrow c_{s,b}^2 + \alpha v_A^2$ where $\alpha$ is an angular factor
depending on the exact velocity and field orientation. To include this in our
work we modify the evolution of the baryon velocity to
\begin{multline}
\label{eq:thetab2}
\dot{\theta}_b = -\clh\, \theta_b + k^2 \, \Psi  + R \tau_c^{-1} \left(\theta_\gamma - \theta_b\right) \\
\mbox{}+ \frac{1}{2} k^2 R \left(\frac{1}{2} \Delta_{B,0} -  \frac{1}{3}
  \Pi_{B,0}^\brsc{0}\right) + k^2 \left( c_{s,b}^2 + \frac{2}{9} v_A^2 \right) \Delta_b
\end{multline}
where $c_{s,b}$ remains the standard unmagnetised baryon sound speed, and we
have defined the Alfv\'en velocity $v_A$ as
\begin{equation}
v_A^2  = \frac{1}{4\pi\rho a^4} \bigl\langle \Bc^2 \bigr\rangle_k
\end{equation}
where $\rho$ is the density of the conducting fluid, and $\langle
\Bc^2\rangle_k$ is the variance of the field from scales larger than $k$. We
discuss this effect in more detail in Appendix~\ref{app:jeans}, motivating the
specific form of the modification in \eqref{eq:thetab2}, in particular the
factor $2/9$ multiplying the Alfv\'en velocity. It is important to note that as
this effect is non-linear, the modes do not decouple and even when evolving the
standard adiabatic mode we must add in the Alfv\'en velocity term.

Radiation free streaming is particularly important for magnetic
fields, and is the most important source of damping on large scales
\cite{Jedamzik:1998}. Prior to free-streaming, there are many
photon-baryon scatterings per wavelength, and the radiation and
baryons appear like a single tightly-coupled conducting
fluid. However, when the photons start to free stream they decouple
from the baryon fluid and this becomes the sole conducting
fluid. Whilst the baryons are no longer tightly coupled to the
photons, there is still enough scattering to exert a significant drag
force on the fluid, and this causes damping on propagating magnetic
waves.

When the photons decouple from a particular scale, the fluid no longer
feels the radiation pressure, and the magnetic pressure dominates the
baryon pressure for $B \gtrsim 0.1 \: \mathrm{nG}$. In this regime,
in addition to Alfv\'en modes, magneto-acoustic modes are also
significant.

Properly accounting for the damping in the radiation free-streaming
regime requires the use of the full Boltzmann system for the photons,
a procedure that is complicated by the necessity of including
non-linear magnetic effects. Instead we use the prescription of
\cite{Subramanian:1998}, who analyse the evolution of Alfv\'en modes
in the presence of a homogeneous radiation drag force. They find that
the magnetic field on small scales damped approximately as
\begin{equation}
\label{eq:magdamp}
\Bc^i(\vk) = \Bc_0^i(\vk) \exp\left(-k^2 \int^{\tau_*} \!\! v_A^2 \tau_c \, d\tau\right) \; ,
\end{equation}
where $v_A$ and $\tau_c$ are defined above, and $\tau_*$ is the time
of last scattering.  For magneto-acoustic modes, the damping is
similar to that of \eqref{eq:magdamp} with the Alfv\'en velocity $v_A$
replaced with the baryon sound speed $c_{s,b}$, causing them to damp
on smaller scales \cite{Jedamzik:1998,Subramanian:1998}. For
simplicity, we will treat the field as damping solely on the largest
scale, that of the Alfv\'en modes.

To account for this damping in our work, we allow the linear
perturbations $\Delta_B$ and $\Pi_B$ to evolve. This is in contrast to
other work in the literature
\cite{Paoletti:2008,Finelli:2008,Yamazaki:2006} where the
perturbations are constant for all time and the damping is imposed as
a cut off on the initial power spectrum. This is implemented by
constructing effective differential equations for the two
perturbations for $\tau < \tau_*$
\begin{align}
\label{eq:magmod1}
\dot{\Delta}_B & = -2 k^2\: v_A^2 \, \tau_c\: \Delta_B \; , \notag \\
\dot{\Pi}_B & = - 2 k^2\: v_A^2  \, \tau_c\: \Pi_B \; .
\end{align}
For $\tau > \tau_*$, we set $\dot{\Delta}_B$ and $\dot{\Pi}_B$ to
zero, as after recombination there is less than one scattering per
Hubble time and the damping is negligible. To ensure that energy
conservation is maintained we presume that this is transferred equally
into the photon-baryon fluid. We modify the baryon density equation to
\begin{equation}
\label{eq:magmod2}
\dot{\Delta}_b = -\theta_b + 3 \dot{\Phi} + \frac{2 k^2}{1 + R^{-1}}  \: v_A^2  \, \tau_c\: \Delta_B \, ,
\end{equation}
and the photon density equation to
\begin{equation}
\label{eq:magmod3}
\dot{\Delta}_\gamma =-\frac{4}{3}\, \theta_\gamma + 4 \dot{\Phi} + \frac{2 k^2}{1 + R^{-1}}  \: v_A^2 \tau_ c\: \Delta_B \, ,
\end{equation}
where we have added the final term in both of the above
equations. Practically, ensuring energy conservation makes little
difference to the results.

There is further avenue of decay for the magnetic fields, through the damping of
MHD turbulence in the post-recombination universe \cite{Sethi:2004}. This effect
is able to damp scales larger than those affected by free-streaming, but smaller
than the Jeans length, though its magnitude is uncertain. This extra damping is
not included in this work, however, as this only modifies modes smaller than the
Jeans length we do not expect it to change our constraints (which are based on
data probing larger scales).

One might worry that the modification to the evolution equations above
invalidate the initial conditions for the magnetic modes that we are using
\cite{Shaw:2009}, which were derived for the unmodified equations. First, the
modification to \eqref{eq:thetab2} takes the form of a modification to the
baryon sound speed, as this is higher order in the initial conditions we do not
expect it to affect our results. The further changes made are to
\eqref{eq:magmod1}, \eqref{eq:magmod2} and \eqref{eq:magmod3}, and are all
proportional to $k^2 \tau \tau_c$ which is negligible on super horizon scales
during radiation domination.

\section{Magnetic Matter Power Spectrum}

Using the above modifications to the evolution equations we are able to
calculate matter power spectra including approximate treatments of the important
non-linear effects. The remaining input is the statistics of the initial
magnetic perturbations $\Delta_B$ and $\Pi_B$. We assume that at some early time
the comoving magnetic field power spectrum is described by a power spectrum
\begin{equation}
\left\langle \Bc_i(\vk) \Bc_j^*(\vk') \right\rangle = (2\pi)^3 \delta(\vk-\vk') \frac{P_{ij}(\khat)}{2} P_B(k) \; ,
\end{equation}
where $P_{ij} = \delta_{ij} - \khat_i \khat_j$, and we will not consider
helicity. The magnetic field power spectrum is
\begin{equation}
P_B(k) = A k^{n_B} \;.
\end{equation}
We will use the spectral index $n_B$ as defined, but it is conventional to give
the amplitude of the spectrum in terms of the variance $B_\lambda^2$ of the
magnetic field strength at a scale $\lambda$ (we choose $\lambda = 1\: \Mpc$ to
agree with the bulk of the literature). This gives
\begin{equation}
  A = \frac{(2\pi)^{n_B + 5} B_\lambda^2}{2\Gamma\left(\frac{n_B+3}{2}\right) k_\lambda^{n_B + 3}} \, .
\end{equation}
To calculate the two required power spectra $P_{\Delta_B}(k)$, $P_{\Pi_B}(k)$
and the cross spectrum $P_{\Delta_B\Pi_B}(k)$ requires convolutions of the
magnetic power spectrum. We use the results of a previous paper
\cite{Shaw:2009}, which allows us to reduce the convolution into a dimensionless
integral depending only on $n_B$. We numerically evaluate these at a large
number of values such that we can accurately interpolate to find power spectra
at an arbitrary $n_B$. It is important to note that as we are including the
damping and Jeans' effects in the evolution of the magnetic perturbations we can
safely avoid imposing a cutoff in the initial power spectrum as used in other
work \cite{Finelli:2008,Paoletti:2010}.

\begin{figure}[tb]
\hspace{-0.04\linewidth}\includegraphics[width=1.04\linewidth]{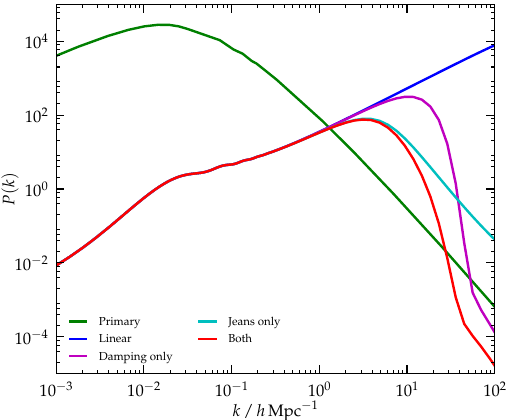}
\caption{The additional non-linear effects modify the magnetic power
  spectrum. We plot: the linear only behaviour; the diffusion damping of the
  magnetic field; the magnetic pressure support; and both effects combined. For
  comparison we also include the primary adiabatic mode. All these
    plots are calculated with a magnetic field strength of $B_\lambda = 5
    \:\mathrm{nG}$ and spectral index $n_B = -2.9$.}
\label{fig:comp_effect}

\end{figure}

In Fig.~\ref{fig:comp_effect} we show the consequence of the two non-linear
effects on the matter power spectrum. The first thing to note is that the linear
theory magnetic power spectrum grows as $P(k) \propto k$. On sub-horizon scales
during matter domination, we can combine the evolution equations for baryons and
dark matter to give an equation for the total matter perturbation $\Delta_m =
R_b \Delta_b + R_c \Delta_c$
\begin{equation}
\ddot{\Delta}_m + \clh \dot{\Delta}_m - \frac{3}{2} \clh^2 \Delta_m = \frac{\rho_\gamma}{\rho_m} k^2 L_B,
\end{equation}
where the Lorentz force $L_B = \frac{2}{3} \left(\frac{1}{3} \Pi_B -
  \Delta_B\right)$, and we have also neglected pressure terms in $c_{s,b}^2$. In
these limits the equation has a simple solution for the magnetic mode
\begin{equation}
\Delta_m = R_b L_B \frac{\rho_\gamma(\tau_i)}{\rho_b(\tau_i)} \left(k \tau_i\right)^2 \Bigl[\frac{1}{10} \Bigl(\frac{\tau}{\tau_i}\Bigr)^2 + \dotsb \Bigr]
\end{equation}
where we have included only the leading order term, and $\tau_i$ is the time
when the mode starts to grow significantly. This is the time that the baryon
perturbation decouples from the photons. Provided this time is similar across a
range of scales (correct for larger scales), we can expect $P_m(k) \propto k^4
P_{L_B}(k)$; for nearly scale invariant magnetic field spectra this gives
$P_m(k) \propto k$.

The effect of including the magnetic pressure increases the Jeans length, and
thus causes smaller scales to oscillate. In matter domination, the comoving
magnetic Jeans wavenumber is constant and thus larger $k$ grow very little after
recombination. On even larger scales the diffusion damping of the magnetic
fields causes the source for the later growth of the magnetic fields to be
exponentially suppressed. This leads to much slower growth in the matter
perturbations, and an effective cutoff beyond which there are no significant
perturbations sourced. The scales at which these effects start roughly agree
with the estimates of \cite{Subramanian:1998,Kim:1996,Tashiro:2009}, where both
are expected to scale like $k_c \propto B_\lambda^{-(n+5)/2}$.

\begin{figure}[tb]
\hspace{-0.04\linewidth}\includegraphics[width=1.04\linewidth]{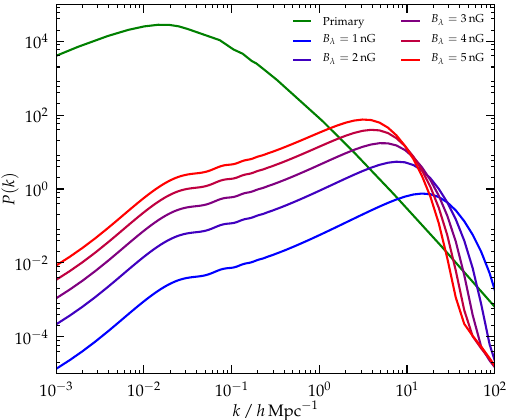}
\caption{The full matter power spectrum with all non-linear effects at a variety
  of magnetic field strengths between $B_\lambda = 1$--$5 \mathrm{nG}$. We have
  held the magnetic spectral index constant at $n_B = -2.9$. The power spectrum
  amplitude increases strongly with that of the magnetic field, however the two
  scales associated with the diffusion damping and magnetic Jeans length also
  increase rapidly.}
\label{fig:comp_amp}

\end{figure}

\begin{figure}[tb]
\hspace{-0.04\linewidth}\includegraphics[width=1.04\linewidth]{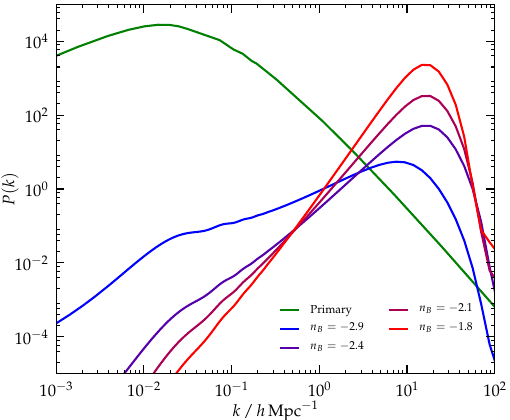}
\caption{The full matter power spectrum with all non-linear effects at a variety
  of magnetic spectral indices between $n_B = -2.9$ to $-1.8$. The
    amplitude is fixed at $B_\lambda = 2\:\mathrm{nG}$.}
\label{fig:comp_ind}

\end{figure}

In Fig.~\ref{fig:comp_amp} and Fig.~\ref{fig:comp_ind}, we plot the effect on
the matter power spectrum of the amplitude and tilt of the magnetic
spectrum. Fig.~\ref{fig:comp_amp} shows that the relative contrast between the
peak magnetic and primary contributions stay roughly constant as the amplitude
$B_\lambda$ is varied. This occurs as a consequence of the fact that the matter
power spectrum at the damping cutoff is $\propto B_\lambda^4 k_c$, whilst the
primary spectrum $\propto k_c^{-3}$. However as the damping wavenumber $k_c
\propto B_\lambda^{-1}$ for nearly scale invariant spectra, the ratio between
the two is constant.

Both the magnetic damping and Jeans effects give small changes to the CMB power
spectra at very high $l$. The maximum scale that is affected by the magnetic
damping is around $k_D \sim 1 \:h\, \mathrm{Mpc}^{-1}$ and this changes the CMB
power spectra on scales $l > D_A k_D \sim 10^4$. For $l < 10^4$ the power
spectra are essentially the same as those shown in \cite{Shaw:2009}, though in
this work our power spectra are evolved from the modified equations.

\section{Results}

We have used versions of \textsc{CAMB} \cite{CAMB} and \textsc{CosmoMC}
\cite{COSMOMC} modified to generate the magnetic contributions to both the CMB
angular power spectrum and the matter power spectrum\footnote{The modified
  version of \textsc{CAMB} and the adaptation of \textsc{CosmoMC} to use it will
  be available from \url{http://camb.info/jrs}.}. We limit ourselves to the most
important magnetic contributions in this work, the magnetic scalar and vector
modes and the tensor passive mode. We sample with a flat prior in the standard
set of cosmological parameters $\left\{ \Omega_b h^2, \Omega_c h^2, \theta,
  \tau, n_s, \log{(10^{10}A_s)} \right\}$, assuming a flat $\Lambda$CDM
universe. To describe the magnetic fields we supplement this with three
parameters, the magnetic power spectrum amplitude $B_\lambda$ and spectral index
$n_B$. We also need to specify the production time of the magnetic fields
$\tau_B$ to calculate the passive modes. The tensor passive mode that we
calculate has the same structure as the standard tensor mode with amplitude
\begin{equation}
H^\brsc{2} = R_\gamma \Pi_B^\brsc{2}\left[ \log{(\tau_\nu / \tau_B)} +
  \frac{5}{8 R_\gamma} - 1\right] \; ,
\end{equation}
where $\tau_\nu$ is the time of neutrino decoupling (see \cite{Shaw:2009}).
Rather than directly specifying the magnetic field production time we use a
proxy $r_B = \log_{10}(\tau_\nu / \tau_B)$, such that $H^\brsc{2} \sim
\log{(10)} \,R_\gamma \Pi_B^\brsc{2} r_B$.

For priors on the magnetic quantities we choose a flat prior in $B_\lambda <
10 \: \mathrm{nG}$. For $r_B$ we use a flat prior with bounds $6 \le r_B \le
17$ corresponding to magnetic field production between the electroweak phase
transition, and reheating at the GUT scale. Finally for the spectral index we
use a flat prior $-2.95 \le n_B \le -1.6$, the lower bound of this comes from
the fact that there is an infrared divergence for $n_B \le -3$. The upper
bound is from the fact that we are primarily interested in primordial fields
produced in the early Universe (prior to nucleosynthesis), the work of
\cite{Caprini:2002} suggests that larger spectral indices are ruled out by
nucleosynthesis constraints (see \S~\ref{sec:caprini}). Other than the upper
bound on the spectral index, our results should be valid for fields produced
at later epochs.

\subsection{CMB only}

Primordial magnetic fields contribute both passive and compensated scalar and
tensor modes, as well as a compensated vector mode. However, the significant
contributions are from the passive tensor mode, and the compensated vector mode,
which are important on large and small scales respectively \cite{Shaw:2009}. For
comparison to CMB data (for scales $l \lesssim 3000$) we need only need
calculate these two.

We use the recent WMAP 7 year release \cite{WMAP7} along with the final data
from QUaD \cite{QUAD} and ACBAR \cite{ACBAR} for $l < 3000$.  We also use the
Hubble Key Project data \cite{HubbleKey}, BBN data, Union Supernova dataset
\cite{SupernovaUnion} and BAO data from SDSS DR7 \cite{BAO}.  To account for the
Sunyaev--Zel'dovich effect at high multipoles we adopt the standard treatment
and use the WMAP template \cite{WMAP7}. This requires an extra parameter for its
normalisation $A_{SZ}$ which we treat as a nuisance parameter and marginalise
over.

\begin{figure}[tpb]
\includegraphics[width=\linewidth]{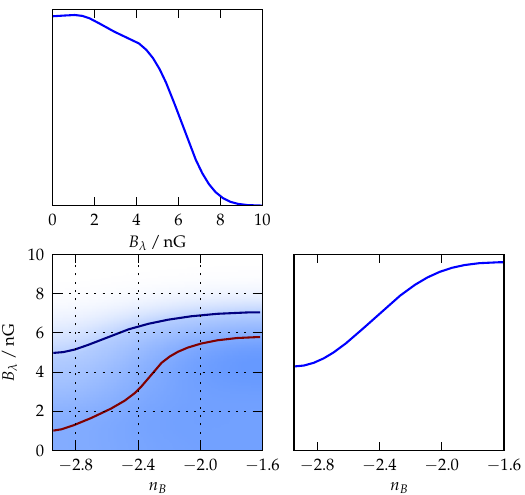}
\caption{The CMB only (WMAP7 + QUAD + ACBAR) constraints on the magnetic field
  amplitude and spectral index. The top left plot is the marginalised constraint
  solely on the magnetic field, and the bottom right is the marginal constraint
  on its spectral index. The bottom left plot is the joint constraint
  (marginalised over the other cosmological parameters), with contours for one
  and two sigma errors.}
\label{fig:cmbonly}
\end{figure}

\begin{figure}[tb]
\includegraphics[width=\linewidth]{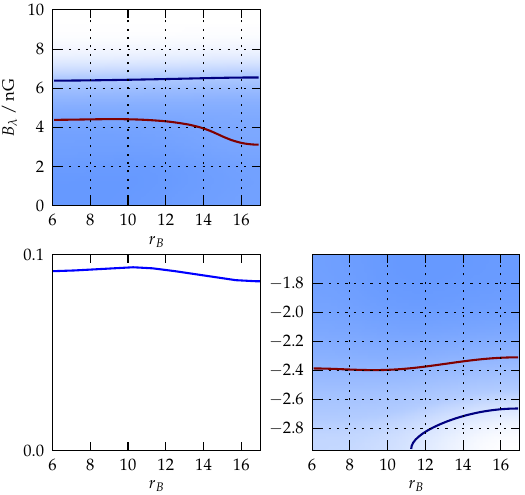}
\caption{The CMB-only marginal constraint on the production time ratio $r_B =
  \log_{10}(\tau_\nu / \tau_B)$ (bottom left), along with the correlations with
  the magnetic field amplitude (top left) and spectral index (bottom
  right). Even in the CMB-only constraint $r_B$ is largely unconstrained and has
  minimal correlation with the other parameters, being only important in the
  case of very large magnetic fields, and a very red spectral index.}

\label{fig:lrat}
\end{figure}

The resulting constraints are shown in Fig.~\ref{fig:cmbonly}, which are in
broad agreement with those of Ref.~\cite{Paoletti:2010}. We have included only
the magnetic amplitude $B_\lambda$ and spectral index $n_B$, marginalising over
the other parameters. The standard cosmological parameters are not shown as they
are in agreement with their values in a universe with no primordial magnetic
fields. We have also chosen not to include $r_B$ as it is unconstrained by the
data (see Fig.~\ref{fig:lrat}). This is a manifestation of the fact that the
tensor passive mode can be only significant at large magnetic amplitudes, and
these are already excluded by the magnetic vector mode. From this we calculate a
95\% Confidence Limit of $B_\lambda < 6.4 \: \mathrm{nG}$.

The constraints on the magnetic parameter space are unchanged when adding in
large-scale matter power data sets such as 2dF \cite{2dF} and SDSS LRG data from
DR4 \cite{SDSS4}\footnote{We do not use the latest SDSS DR7 data \cite{SDSS7}
  due to complications modifying its likelihood calculation to include the
  magnetic field matter power spectrum.}, though the cosmological parameters
change as expected. The reason can be seen from Fig.~\ref{fig:comp_amp}, as the
magnetic contributions are only significant for $k \gtrsim 1\: h\, \Mpc^{-1}$,
and the galaxy redshift surveys probe only up to $k \sim 0.2 \: h \:
\Mpc^{-1}$. We will include this data for the remainder of this paper.

\subsection{Sunyaev---Zel'dovich Effect}

Recent work \cite{Tashiro:2009} has suggested that the Sunyaev--Zel'dovich (SZ)
effect may be able to give tight constraints on the magnetic power spectrum. As
the magnetic fields cause an increase in the small scale matter power spectrum,
this gives rise to a large number of small mass halos and thus a rise in the SZ
angular power spectrum. The recent release of data from the South Pole Telescope
(SPT) \cite{SPT}, provides CMB observations up to $l \sim 10^4$ that we will
compare to.

We follow the general method of \cite{Tashiro:2009}, using the prescription of
\cite{Komatsu:2002} to calculate the thermal SZ angular power spectrum from the
matter power spectrum, which we have calculated using our modified version of
CAMB. We give the details of this calculation in Appendix~\ref{app:sz}. In
addition to the SZ contribution the magnetic vector mode gives a significant
addition to the small scale CMB power, which must be included. We do not include
the kinetic SZ spectrum which is sub-dominant to the thermal spectrum. We also
note that the non-gaussian statistics of the SZ effect can significantly
increase its intrinsic variance \cite{Zhang:2007}; we do not account for this
within this paper.

\begin{figure}
\includegraphics[width=\linewidth]{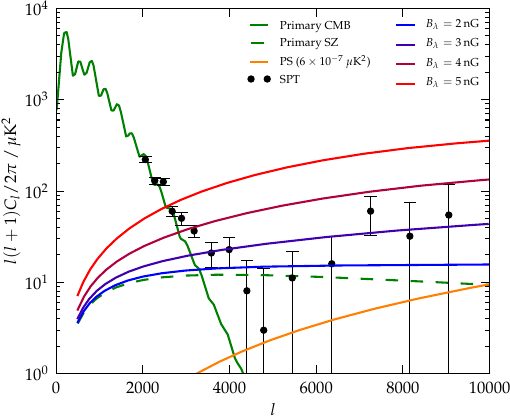}
\caption{The angular power spectrum of the CMB and the SZ contributions to
  it. We plot the SZ contribution from the primary adiabatic mode only, and in
  combination with four magnetic field strengths $B_\lambda =
  2$--$5\:\mathrm{nG}$, with a magnetic spectral index of $n_B = -2.4$. We also
  plot the bandpowers from SPT, and the estimate of the residual point source
  contribution (labelled PS). Both the SZ contribution and the SPT data are at
  $150 \:\mathrm{GHz}$.}
\label{fig:sz}
\end{figure}
As well as the datasets for the CMB-only constraints, and the SPT data, we have
also included the large scale matter power data. When generating the SZ power
spectrum we must include all contributions, both the adiabatic and magnetic
contributions, as such we do not add in an SZ template, and thus $A_{SZ}$ is no
longer sampled over. The SPT data contains a further small scale contribution
from the unresolved point sources, both from star-forming and radio galaxies
(see Fig.~\ref{fig:sz}). The exact value must be fit from the bandpowers (in the
analysis of \cite{SPT} it is poorly constrained with $C_l^\text{ps} = (6.2 \pm
6.4) \times 10^{-7}\:\mu\mathrm{K}^2$). We expect that neglecting this
contribution will have resulted in a small increase to our upper limits.

The magnetic contribution to the SZ power spectrum from our calculations is
smaller than that of \cite{Tashiro:2009}. Differences between our calculations,
such as in finding the amplitude of the matter perturbations and treating the
non-linear effects, make it difficult to give a single reason for this.

\begin{figure}[tpb]
\includegraphics[width=\linewidth]{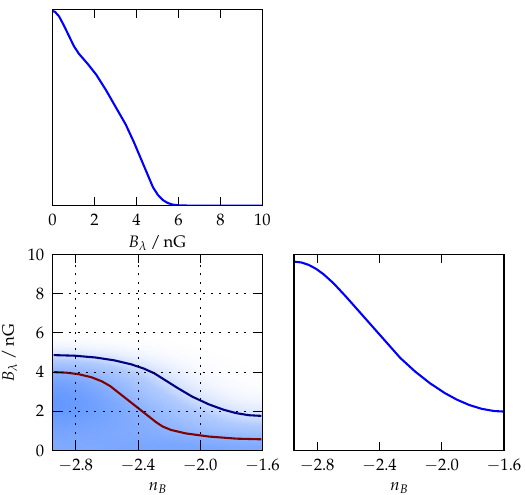}
\caption{The CMB and SZ constraints on the magnetic field amplitude and spectral
  index. The individual plots are equivalent to those in
  Fig.~\ref{fig:cmbonly}. Calculating the magnetic SZ contributions with the
  addition of SPT data has restricted the parameter space to lower amplitudes
  and redder spectral indices than the large scale CMB data.}
\label{fig:cmbsz_nolya}
\end{figure}

Figure~\ref{fig:cmbsz_nolya} shows the marginalised probability distributions
for the magnetic parameters (again we do not plot $r_B$). The addition of the
SPT data has excluded much of the parameter space that was allowed when
including primary CMB effects only, especially the region with large magnetic
field and blue spectral index, which gives the most dramatic change in the
matter power spectrum (see Fig.~\ref{fig:comp_ind}). This gives a large decrease
in the 95\% CL of the magnetic amplitude to $B_\lambda < 4.1 \: \mathrm{nG}$.

\subsection{Lyman Alpha Data}
\label{sec:lya}
It is clear from both Fig.~\ref{fig:comp_amp} and Fig.~\ref{fig:comp_ind} that
the small scale matter power spectrum is significantly affected by the presence
of a primordial magnetic field. Unfortunately galaxy redshift surveys such as
SDSS probe only as far as $k \lesssim 0.2 \: h \: \Mpc^{-1}$, too large scales
to be affected by the magnetic fields. However, observations of the
Lyman-$\alpha$ flux power spectrum probe the matter density power spectrum to
scales as small as $k \sim 5 \: h \: \Mpc^{-1}$, and may be able to give a much
more powerful constraint.  Unfortunately there is no simple analytic mapping
from the observations to the matter spectrum, so cosmological constraints need
to be obtained by comparison to simulations \cite{McDonald:2005}. A fully
consistent analysis including magnetic fields is beyond the scope of this paper;
instead we use a very rough simple approximation to the likelihood.

We use the standard Lyman-$\alpha$ module in \textsc{CosmoMC} which is based on
the work of \cite{McDonald:2005}. This finds an effective amplitude and spectral
index about a pivot scale of $k = 0.009\: \mathrm{s}\, \mathrm{km}^{-1}$
(roughly $k \sim 1 \: h \: \Mpc^{-1}$). It calculates a likelihood from these by
interpolating between a set of simulations compared to SDSS quasar data. Due to
the large difference between our power spectra (when including magnetic effects)
and those of $\Lambda$CDM we cannot expect the likelihoods to be very accurate,
especially at large magnetic amplitudes and blue spectral indices, where the
effect is greatest. However for the pivot scale, and range of scales probed by
the SDSS spectra used $k \lesssim 0.02\: \mathrm{s}\, \mathrm{km}^{-1}$, the
magnetic contribution is generally small compared to that of the primary
adiabatic mode. In this light our results should be viewed as an approximation
to the constraints that a more sophisticated likelihood approach would achieve.

\begin{figure}[tpb]
\includegraphics[width=\linewidth]{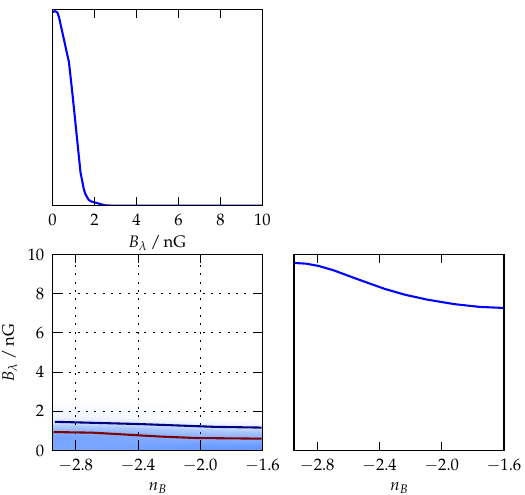}
\caption{Constraints on the magnetic field amplitude and spectral index using
  CMB and matter power data, including Lyman-$\alpha$ data. The Lyman-$\alpha$
  data, which probes the small scale matter distribution, has dramatically
  reduced the allowable parameter space, to a range of small amplitudes
  ($B_\lambda \lesssim 1.5 \: \mathrm{nG}$) roughly independent of spectral
  index $n_B$. }
\label{fig:cmb_lya}
\end{figure}

The magnetic parameter space is significantly constricted by the use of the
Lyman-$\alpha$ data (see Fig.~\ref{fig:cmb_lya}), with the allowed region for
$B_\lambda$ being largely independent of the spectral index $n_B$. This is to
be expected as the Lyman-$\alpha$ pivot ($k \sim 1 \: h \:\Mpc^{-1}$)
coincides with the scale at which the magnetic matter power spectrum amplitude
is similar across a broad range of $n_B$. Overall this results in a 95\% CL of
$B_\lambda < 1.3 \: \mathrm{nG}$.


\subsection{Nucleosynthesis Constraints}

\label{sec:caprini}

Nucleosynthesis places strong constraints on the amount of energy density in
gravitational waves allowed in the Universe, giving a limit of
$\Omega_\text{GW} \lesssim 1.1 \times 10^{-6}$ \cite{Maggiore:2000}.
\changeR{Prior to neutrino decoupling the anisotropic stress of an
inhomogeneous magnetic field on superhorizon scales sources gravitational
waves\footnote{These are the same as tensor passive modes \cite{Shaw:2009},
though they stop growing when they enter the horizon prior to neutrino
decouping.}, this allows the small scale magnetic fields to transfer some of
their energy into gravitational waves before it is dissipated into the photon-
baryon plasma \cite{Caprini:2002}. This process is quite efficient, and gives
the gravitational waves a significant fraction of the original magnetic energy
(see \cite{Caprini:2005} for an intuitive explanation of this).}  As the
magnetic field energy density is necessarily blue, there is more energy
density in the smallest scale fluctuations. This allows us to strengthen the
usual constraints on the total magnetic energy density which only take into
account the energy \changeR{remaining} in the field at nucleosynthesis by
constraining $\Omega_\text{GW}$ which is sourced at a time when much smaller
scales have not been damped. We take Eq.~(33) from \cite{Caprini:2002} adapted
to our conventions, yielding

\begin{equation}
B_\lambda \: / \: \mathrm{nG} < 700\,h\, \left[2^{\frac{n+5}{2}} \Gamma\left({\textstyle \frac{n+5}{2}}\right)\right]^{1/2} 10^{-\frac{n+3}{2} \left(4 + r_B\right)} \; .
\end{equation}
It is essential to note that this assumes that the magnetic field is well
described by a single power law across a vast range of scales --- from the
pivot scale at $k = 1\: \Mpc^{-1}$ up to at least $k = 10^{10} \Mpc^{-1}$, the
horizon scale at the electroweak phase transition. Depending on how the
magnetic field is generated this assumption may break down---any freedom to
reduce small scale power would significantly weaken the constraints.

\begin{figure}[tpb]
\includegraphics[width=\linewidth]{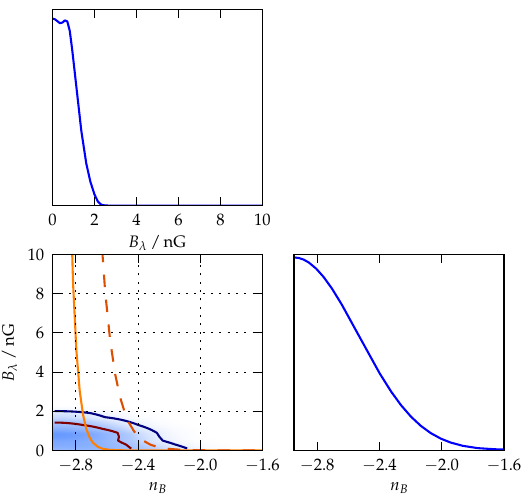}
\caption{Constraints on the magnetic parameter space from Lyman-$\alpha$ data
  including the nucleosynthesis constraints of \cite{Caprini:2002}. The solid
  and dashed orange lines give the upper bounds in amplitude of a magnetic field
  produced at GUT scale inflation and the electroweak phase transition
  respectively. This corresponds to our priors.}
\label{fig:cmb_gw}
\end{figure}

In Fig.~\ref{fig:cmb_gw} we show the probability distribution of the magnetic
parameters when combining the nucleosynthesis constraint with the Lyman-$\alpha$
data of \S~\ref{sec:lya}. As we would expect this reduces the allowed parameter
spaces to redder spectral indices $n_B \lesssim -2$. Because of this the limits
on the amplitude are slightly enlarged with the 95\% CL becoming $B_\lambda <
1.6 \: \mathrm{nG}$.



\section{Conclusion}

As we have seen in this paper, when considering the magnetic
contributions to the matter power spectrum it is essential to treat
important small-scale, non-linear effects. We have demonstrated a
technique for approximating the main non-linear effects within linear
perturbation theory, and have incorporated this into a modified
version of \textsc{CAMB}. This gives an alternative to the common
approach of incorporating these effects directly into the initial
power spectrum, such as in Ref.~\cite{Paoletti:2010}.

We have used our theoretical predictions to place constraints on the allowable magnetic field
amplitude given various data sets. Where all limits are at 95\%
confidence, using CMB data only we find $B_\lambda < 6.4 \:
\mathrm{nG}$, with a redder spectral index favoured. As the presence
of a stochastic magnetic field gives significant modifications to the
small scale matter distribution, we also look at the constraints when
adding two probes of it: the Sunyaev--Zel'dovich effect measured by
SPT \cite{SPT}, which gives a limit of $B_\lambda < 4.1 \:
\mathrm{nG}$; and Lyman-$\alpha$ forest data from SDSS
\cite{McDonald:2005}, which gives a rough constraint of $B_\lambda \alt 1.3 \:
\mathrm{nG}$.

Whilst the addition of the small-scale matter data gives a large
reduction in the allowed amplitude of a primordial magnetic field
(certainly when using Lyman-$\alpha$ data), this is still roughly
consistent with a scenario where current galactic fields are formed
solely by adiabatic contraction of primordial fields. 
Increasing the accuracy of measurements at the scales we
can currently probe with data will provide limited gains: since the
magnetic power spectrum increases with $B_\lambda^4$,
significantly decreasing the errors on matter power spectrum
measurements will produce less impressive gains in the upper limit on
$B_\lambda$. However, if better observations and modelling allow
accurate comparison to smaller scales there should be an almost linear
decrease in the upper limit with the smallest scale probed. This would
seem to provide the best opportunity for testing the primordial field
hypothesis.


\nopagebreak[4]\subsection*{Acknowledgements} We would like to thank Steven
Gratton, Simeon Bird, Chiara Caprini and Camille Bonvin for useful discussion.
JRS was supported by an STFC studentship. AL acknowledges a PPARC/STFC advanced
fellowship.


\appendix

\section{Motivating the Magnetic Jeans Effect}
\label{app:jeans}
To analyse the magnetic Jeans effect we need to look at the non-linear evolution
of the field perturbations. First let us construct $\Delta^{ij}$, a quadratic
combination of the magnetic field
\begin{equation}
\Delta^{ij} = \frac{1}{4\pi \rho_\gamma a^4} \Bc^i \Bc^j,
\end{equation}
with the normalisation chosen such that it is conveniently close to the magnetic
perturbations
\begin{align}
\label{eq:pertdef}
\Delta_B &= \frac{1}{2} \Delta^k_k \notag \; ,\\
\Pi_B^{ij} & = \Delta^k_k \delta^{ij} -3 \Delta^{ij} \; .
\end{align}
We are interested in the time evolution of $\Delta^{ij}$, and thus want to
calculate the derivative $\dot{\Delta}^{ij} = \frac{1}{2\pi \rho_\gamma a^4}
\Bc^{(i}\partial_\tau \Bc^{j)}$, where the parentheses indicate symmetrisation
with respect to the enclosed indices. It will be useful to re-express
\eqref{eq:mhd} by expanding the Levi-Civita symbols
\begin{equation}
\label{eq:mhd2}
\partial_\tau \Bc^i = \Bc^k (\partial_k v^i) - v^k (\partial_k \Bc^i) - \Bc^i \partial_k v^k \; .
\end{equation}
Using \eqref{eq:mhd2} we can calculate the time derivative of
$\dot{\Delta}^{ij}$
\begin{equation}
  \dot{\Delta}^{ij} = 2 ( \partial_k v^{(i} ) \Delta^{j)k} - 2 (\partial_k v^k) \Delta^{ij} - v^k (\partial_k \Delta^{ij}) \; .
\end{equation}
Using the definitions of \eqref{eq:pertdef} we can split the above into
equations for each of the perturbations $\Delta_B$ and $\Pi_B$. First
\begin{equation}
\label{eq:dotdelb}
\dot{\Delta}_B = - \frac{4}{3} \Delta_B \theta_b - v^k (\partial_k \Delta_B) - \frac{1}{3} \sigma_{kl} \Pi_B^{kl}
\end{equation}
where we have only included terms up to second order, and we have decomposed
\begin{equation}
\partial_i v_j = \frac{1}{3} \theta_b \delta_{ij} + \sigma_{ij} + \omega_{ij}
\end{equation}
at linear order. Physically $\theta_b$ is the divergence of the baryon
3-velocity, and $\sigma_{ij}$ is its shear; we will neglect the antisymmetric
vorticity tensor $\omega_{ij}$. Second, the magnetic anisotropic stress evolves as
\begin{multline}
\label{eq:dotpib}
\dot{\Pi}_B^{ij} = - \frac{4}{3} \theta_b \Pi_B^{ij} - v^k\partial_k \Pi_B^{ij} - 4 \sigma^{ij} \Delta_B \\+ 2 \sigma_k{}^{(i}\Pi_B^{j)k} - \frac{2}{3} \sigma_{kl}\Pi_B^{kl} \delta^{ij} \; .
\end{multline}

In this paper we do not attempt a fully self-consistent non-linear
analysis. Instead, we try to identify the most important non-linear effects and
then model them using an approximate method that is at least as consistent as
most previous work. Here we are interested in the magnetic Jeans instability ---
the isotropic part of this effect is described by the first terms on the
right-hand side of \eqref{eq:dotdelb} and \eqref{eq:dotpib}:
\begin{align}
\label{eq:neglected}
\dot{\Delta}_B & = - \frac{4}{3} \Delta_B \theta_b \notag \, ,\\
\dot{\Pi}_B^{ij} &= - \frac{4}{3} \Pi_B^{ij} \theta_b  \; ,
\end{align}
corresponding to an enhancement $\propto \left(\rho_b/\rho_b^0\right)^{4/3}$ as
a perturbation collapses to density $\rho_b$ compared to a background value of
$\rho_b^0$. To leading order we can use the linear baryon density evolution
equation
\begin{equation}
\dot{\Delta}_b = - \theta_b + 3 \dot{\Phi}
\end{equation}
to replace $\theta_b$, and neglect $\dot{\Phi}$ on the small scales of interest,
giving
\begin{align}
\label{eq:magpertevo}
\Delta_B & = \Delta_{B,0} \left( 1 + \frac{4}{3}\Delta_b\right)
 \notag \\
\Pi_B^{ij} & = \Pi_{B,0}^{ij} \left( 1 + \frac{4}{3}\Delta_b\right)
 \; ,
\end{align}
where $\Delta_{B,0}$ and $\Pi_{B,0}$ are the magnetic perturbations
at some initial time where the baryon perturbation $\Delta_b$
was much smaller than its present value. The linearised equation for the
evolution of the baryon velocities in Fourier space is
\begin{multline}
\label{eq:thetab}
\dot{\theta}_b = -\clh\, \theta_b + k^2 \, c_{s,b}^2 \Delta_b + k^2 \, \Psi  + R \tau_c^{-1} \left(\theta_\gamma - \theta_b\right) \\
\mbox{}+ \frac{1}{2} k^2 R \left(\frac{1}{2} \Delta_B -  \frac{1}{3} \Pi_B^\brsc{0}\right) \;
\end{multline}
where $\tau_c$ is defined as $\tau_c = 1/(a n_e \sigma_T)$, $c_{s,b}$ the baryon
sound speed, $R = 4\rho_\gamma / 3\rho_b$, and the last term is the magnetic
interaction, the Lorentz force. To include the magnetic field evolution in this,
we use Eq.\eqref{eq:magpertevo}, and replace the magnetic perturbation with its
expectation (smoothed at the relevant scale) in the higher order terms only.
This is not fully justified, however the magnetic pressure is always positive,
and there are large-scale modes that look locally homogeneous, so this
prescription aims to include the main qualitative effect of the pressure
enhancement due to the large-scale magnetic field being adiabatically
compressed. The corresponding effect from large-scale densities compressing the
small-scale field has random sign, and hence is expected to have zero mean. In
Fourier space this leaves
\begin{align}
\label{eq:magpertevo2}
\Delta_B & = \Delta_{B,0} + \frac{4}{3} \bigl\langle \Delta_{B,0} \bigr\rangle_k \Delta_b
\, , \\
\Pi_B^{ij} & = \Pi_{B,0}^{ij} \; ,
\end{align}
where the expectation is evaluated using only modes larger than the scale of
interest.  The equation for $\Pi_B$ has only the lowest order term as $\langle
\Pi_{B,0} \rangle_k = 0$. As a caveat we note that this averaging procedure it
unlikely to be valid if the objective is to study non-linear collapse at density
peaks: the large-scale background field changes from place to place, and the
density of collapsed objects is expected to be correspondingly modulated. For
calculating the power spectrum it may however be a reasonable approximation, and
our later results are in fact most constrained by the power spectrum.

Inserting \eqref{eq:magpertevo2} into \eqref{eq:thetab} gives
\begin{multline}
\label{eq:thetab3}
\dot{\theta}_b = -\clh\, \theta_b + k^2 \, c_{s,b}^2 \Delta_b + k^2 \, \Psi  + R \tau_c^{-1} \left(\theta_\gamma - \theta_b\right) \\
\mbox{}+ \frac{1}{2} k^2 R \left(\frac{1}{2} \Delta_{B,0} -  \frac{1}{3} \Pi_{B,0}^\brsc{0}\right) + \frac{2}{9} k^2 v_A^2 \Delta_b
\end{multline}
where we have defined the Alfv\'en velocity $v_A$ as
\begin{equation}
v_A^2  = \frac{1}{4\pi\rho a^4} \bigl\langle \Bc^2 \bigr\rangle_k = \frac{3}{2} R \bigl\langle \Delta_{B,0} \bigr\rangle_k \; ,
\end{equation}
where $\rho$ is the density of the conducting fluid.  The last equality comes
from the fact that during matter domination $\rho = \rho_b$.
This is the standard evolution equation \eqref{eq:thetab} with a new
effective sound speed $c_s^2 \rightarrow c_s^2 + \frac{2}{9} v_A^2$. This
agrees with other approximate treatments of the magnetic Jeans effect in the
literature up to the factor of $\frac{2}{9}$
\cite{Kim:1996,Adams:1996,Subramanian:1998}, which is different but of the
same order. This discrepancy has little effect on the magnetic Jeans scale,
changing it by at most a factor of $~2$.

\section{Sunyaev---Zel'dovich Effect}
\label{app:sz}
In order to compare to the recent data from the South Pole Telescope
\cite{SPT}, we need to be able to calculate SZ angular power spectra
from linear matter power spectra. We use the halo method of Komatsu
and Seljak \cite{Komatsu:2002}, and largely follow details of the
calculation in \cite{SPT}. We give an outline of this below.

The angular power spectrum is given by
\begin{equation}
C_l = g_\nu^2 \int\! dz\, \frac{d V_c}{dz} \int\! dM\, \frac{d n(M,z)}{dM} \left\lvert y_l(M, z) \right\rvert^2 \; .
\end{equation}
In the above $V_c$ is the comoving volume out to redshift $z$, and $g_\nu$ is the spectral function given by
\begin{equation}
g_\nu = \frac{x}{\tanh(x/2)} - 4
\end{equation}
where $x = h\nu / k_B T_\text{CMB}$. The halo mass function $\frac{d
  n}{dM}$ is comoving number density of virialized halos at mass
$M$. Finally $y_l$ is the Fourier transform of the projected Compton
$y$ profile
\begin{equation}
y_l(M,z) = \frac{4\pi r_s}{l_s^2} \int_0^\infty y_{3D}(x) \sinc{(l x / l_s)} x^2 dx \; ,
\end{equation}
where $r_s$ is the scale radius of the profile, $l_s = d_a / r_s$ is
its angular projection ($d_a$ is the angular diameter distance to
redshift $z$) and $y_{3D}$ is the three-dimensional Compton profile in
terms of $x = r / r_s$. The profile $y_{3D}$ is determined by the
model chosen for the baryon density and temperature profile of the
halo. To find it we follow the details of \cite{Komatsu:2002}, fixing
its form with four assumptions: the dark matter density profile is NFW
\cite{NFW}; hydrostatic equilibrium between the gas pressure and the
halo self-gravity; baryon density traces the dark matter density in
the outer halo; and the gas has a polytropic equation of state $P_b
\propto \rho_b^\gamma$. The results of this are given below, for
details see \cite{Komatsu:2002}.

Using an NFW dark matter profile, the scale radius above $r_s$ is the
usual NFW definition, $r_s = r_\text{vir} / c$ where $c$ is the
concentration, and $r_\text{vir}$ is the virial radius given by
\begin{equation}
r_\text{vir} = \left(\frac{3}{4\pi} \frac{M}{\Delta_c \rho_\text{cr}}\right)^{1/3} \; ,
\end{equation}
where the virialisation parameter $\Delta_c$ can be calculated from
the spherical collapse of a top hat perturbation. A fitting formula
for $\Delta_c$ is calculated in \cite{Bryan:1998}
\begin{equation}
\Delta_c(z) = 18\pi^2 - 82 \Omega_\Lambda\left(z\right) + 39 \Omega_\Lambda\left(z\right)^2
\end{equation}
which is accurate in the range $\Omega_\Lambda < 0.9$.

The concentration parameter $c$ which defines the scale radius of the
profile can be fitted from simulations. We use the relation of
\cite{Duffy:2008} which takes the form
\begin{equation}
c(M,z) \approx \frac{7.85}{(1+z)^{0.71}} \left(\frac{M}{M_*}\right)^{-0.081} \; ,
\end{equation}
where the pivot mass is fixed to be $M_* = 2 \times 10^{12}\, h^{-1}\: \msolar$.

The profle $y_{3D}$ is given by
\begin{multline}
y_{3D}(x) = 1.14\times10^{-4} \,\Mpc^{-1} \\ \times\left[\frac{\rho_b(0)}{10^{14} \msolar \Mpc^{-3}}\right] \left[\frac{k_B T_b(0)}{8\, \mathrm{keV}}\right] y(x) \; .
\end{multline}
The dimensionless function $y(x)$ gives the profile shape
\begin{equation}
y(x) = \Bigl[1 - B\bigl[1 - x^{-1} \ln{(1+x)}\bigr]\Bigr]^{\gamma/(\gamma - 1)} \; .
\end{equation}
$B$ is a constant given by
\begin{equation}
B \equiv 3 \eta^{-1} \frac{\gamma - 1}{\gamma} \left[\frac{1}{c}\ln{(1+c)} - \frac{1}{1+c}\right]^{-1}
\end{equation}
Fitting functions for $\eta$ and $\gamma$ are derived in
\cite{Komatsu:2002}. They are valid for the range $1 < c < 25$:
\begin{align}
\gamma & = 1.137 + 0.0894 \, \ln{(c/5)} - 3.68\times 10^{-3} (c-5) \; ,\\
\eta & = 2.235 + 0.202\, (c-5) - 1.16 \times 10^{-3}  (c-5)^2 \; .
\end{align}
The central gas density $\rho_b(0)$ is
\begin{multline}
\rho_b(0) = 7.96\times 10^{12} \msolar \,\Mpc^{-3} \;  \left(\frac{M}{10^{14} \msolar}\right)\left(\frac{r_\text{vir}}{\Mpc}\right)^3 \\
\times \left(\frac{\Omega_b}{\Omega_m}\right) \frac{y(x)^{-1/\gamma}}{(1+c)^2} \left[\frac{1}{c}\ln{(1+c)} - \frac{1}{1+c}\right]^{-1} \; ,
\end{multline}
and the central temperature $T_b(0)$ is
\begin{equation}
T_b(0) = 0.880\, \mathrm{keV} \, \eta \left(\frac{M}{10^{14} \msolar}\right)\left(\frac{r_\text{vir}}{\Mpc}\right)^{-1} \; .
\end{equation}
The final ingredient needed to calculate the SZ power spectrum is the
mass function. In common with \cite{Komatsu:2002,SPT} we use the
Jenkins mass function \cite{Jenkins} calculated from N-body
simulations. As with the Press-Schechter prescription the key quantity
is the smoothed variance $\sigma(R)$ defined by
\begin{equation}
\sigma^2(R; z) =  \int d\ln k \: \widetilde{W}_R^2(k) \, \mathcal{P}(k; z) \; .
\end{equation}
In our work we choose a window function $\widetilde{W}_R(k)$ that is a
spherical top hat in real space. The mass enclosed in this comoving
scale is simply $M = 4\pi \rho_{m,0} R^3 / 3$, and this defines an
obvious mapping between a mass smoothed $\sigma(M)$ and
$\sigma(R)$. In terms of $\sigma(M)$ the mass function of
\cite{Jenkins} is
\begin{equation}
\frac{M^2}{\rho_{m,0}} \frac{d n(M; z)}{d M} = 0.301 \left\lvert\frac{d \ln{\sigma}}{d \ln{M}} \right\rvert \exp{\Bigl(-\bigl\lvert 0.64 - \ln{\sigma} \bigr\rvert\Bigr)}
\end{equation}


\bibliography{neutrino,antony}

\providecommand{\aj}{Astron. J. }\providecommand{\apj}{Astrophys. J.
  }\providecommand{\apjl}{Astrophys. J.
  }\providecommand{\mnras}{MNRAS}\providecommand{\aap}{Astron. Astrophys.}
\begin{thebibliography}{62}
\expandafter\ifx\csname natexlab\endcsname\relax\def\natexlab#1{#1}\fi
\expandafter\ifx\csname bibnamefont\endcsname\relax
  \def\bibnamefont#1{#1}\fi
\expandafter\ifx\csname bibfnamefont\endcsname\relax
  \def\bibfnamefont#1{#1}\fi
\expandafter\ifx\csname citenamefont\endcsname\relax
  \def\citenamefont#1{#1}\fi
\expandafter\ifx\csname url\endcsname\relax
  \def\url#1{\texttt{#1}}\fi
\expandafter\ifx\csname urlprefix\endcsname\relax\def\urlprefix{URL }\fi
\providecommand{\bibinfo}[2]{#2}
\providecommand{\eprint}[2][]{\url{#2}}

\bibitem[{\citenamefont{{Widrow}}(2002)}]{Widrow:2002}
\bibinfo{author}{\bibfnamefont{L.~M.} \bibnamefont{{Widrow}}},
  \bibinfo{journal}{Reviews of Modern Physics} \textbf{\bibinfo{volume}{74}},
  \bibinfo{pages}{775} (\bibinfo{year}{2002}), \eprint{arXiv:astro-ph/0207240}.

\bibitem[{\citenamefont{{Kulsrud} and {Zweibel}}(2008)}]{Kulsrud:2008}
\bibinfo{author}{\bibfnamefont{R.~M.} \bibnamefont{{Kulsrud}}}
  \bibnamefont{and} \bibinfo{author}{\bibfnamefont{E.~G.}
  \bibnamefont{{Zweibel}}}, \bibinfo{journal}{Reports on Progress in Physics}
  \textbf{\bibinfo{volume}{71}}, \bibinfo{pages}{046901}
  (\bibinfo{year}{2008}), \eprint{0707.2783}.

\bibitem[{\citenamefont{{Hogan}}(1983)}]{Hogan:1983}
\bibinfo{author}{\bibfnamefont{C.~J.} \bibnamefont{{Hogan}}},
  \bibinfo{journal}{Physical Review Letters} \textbf{\bibinfo{volume}{51}},
  \bibinfo{pages}{1488} (\bibinfo{year}{1983}).

\bibitem[{\citenamefont{{Durrer} and {Caprini}}(2003)}]{Durrer:2003}
\bibinfo{author}{\bibfnamefont{R.}~\bibnamefont{{Durrer}}} \bibnamefont{and}
  \bibinfo{author}{\bibfnamefont{C.}~\bibnamefont{{Caprini}}},
  \bibinfo{journal}{Journal of Cosmology and Astro-Particle Physics}
  \textbf{\bibinfo{volume}{11}}, \bibinfo{pages}{10} (\bibinfo{year}{2003}),
  \eprint{arXiv:astro-ph/0305059}.

\bibitem[{\citenamefont{{Demozzi} et~al.}(2009)\citenamefont{{Demozzi},
  {Mukhanov}, and {Rubinstein}}}]{Demozzi:2009}
\bibinfo{author}{\bibfnamefont{V.}~\bibnamefont{{Demozzi}}},
  \bibinfo{author}{\bibfnamefont{V.}~\bibnamefont{{Mukhanov}}},
  \bibnamefont{and}
  \bibinfo{author}{\bibfnamefont{H.}~\bibnamefont{{Rubinstein}}},
  \bibinfo{journal}{\jcap} \textbf{\bibinfo{volume}{8}}, \bibinfo{pages}{25}
  (\bibinfo{year}{2009}), \eprint{0907.1030}.

\bibitem[{\citenamefont{{Bernet} et~al.}(2008)\citenamefont{{Bernet},
  {Miniati}, {Lilly}, {Kronberg}, and {Dessauges-Zavadsky}}}]{Bernet:2008}
\bibinfo{author}{\bibfnamefont{M.~L.} \bibnamefont{{Bernet}}},
  \bibinfo{author}{\bibfnamefont{F.}~\bibnamefont{{Miniati}}},
  \bibinfo{author}{\bibfnamefont{S.~J.} \bibnamefont{{Lilly}}},
  \bibinfo{author}{\bibfnamefont{P.~P.} \bibnamefont{{Kronberg}}},
  \bibnamefont{and}
  \bibinfo{author}{\bibfnamefont{M.}~\bibnamefont{{Dessauges-Zavadsky}}},
  \bibinfo{journal}{\nat} \textbf{\bibinfo{volume}{454}}, \bibinfo{pages}{302}
  (\bibinfo{year}{2008}), \eprint{0807.3347}.

\bibitem[{\citenamefont{{Subramanian} and
  {Barrow}}(1998{\natexlab{a}})}]{Subramanian:1998b}
\bibinfo{author}{\bibfnamefont{K.}~\bibnamefont{{Subramanian}}}
  \bibnamefont{and} \bibinfo{author}{\bibfnamefont{J.~D.}
  \bibnamefont{{Barrow}}}, \bibinfo{journal}{\prl}
  \textbf{\bibinfo{volume}{81}}, \bibinfo{pages}{3575}
  (\bibinfo{year}{1998}{\natexlab{a}}), \eprint{arXiv:astro-ph/9803261}.

\bibitem[{\citenamefont{{Mack} et~al.}(2002)\citenamefont{{Mack},
  {Kahniashvili}, and {Kosowsky}}}]{Mack:2002}
\bibinfo{author}{\bibfnamefont{A.}~\bibnamefont{{Mack}}},
  \bibinfo{author}{\bibfnamefont{T.}~\bibnamefont{{Kahniashvili}}},
  \bibnamefont{and}
  \bibinfo{author}{\bibfnamefont{A.}~\bibnamefont{{Kosowsky}}},
  \bibinfo{journal}{\prd} \textbf{\bibinfo{volume}{65}},
  \bibinfo{pages}{123004} (\bibinfo{year}{2002}),
  \eprint{arXiv:astro-ph/0105504}.

\bibitem[{\citenamefont{{Kahniashvili} and {Ratra}}(2005)}]{Kahniashvili:2005}
\bibinfo{author}{\bibfnamefont{T.}~\bibnamefont{{Kahniashvili}}}
  \bibnamefont{and} \bibinfo{author}{\bibfnamefont{B.}~\bibnamefont{{Ratra}}},
  \bibinfo{journal}{\prd} \textbf{\bibinfo{volume}{71}}, \bibinfo{eid}{103006}
  (\bibinfo{year}{2005}), \eprint{arXiv:astro-ph/0503709}.

\bibitem[{\citenamefont{{Gopal} and {Sethi}}(2005)}]{Gopal:2005}
\bibinfo{author}{\bibfnamefont{R.}~\bibnamefont{{Gopal}}} \bibnamefont{and}
  \bibinfo{author}{\bibfnamefont{S.~K.} \bibnamefont{{Sethi}}},
  \bibinfo{journal}{\prd} \textbf{\bibinfo{volume}{72}}, \bibinfo{eid}{103003}
  (\bibinfo{year}{2005}), \eprint{arXiv:astro-ph/0506642}.

\bibitem[{\citenamefont{{Kahniashvili}
  et~al.}(2009)\citenamefont{{Kahniashvili}, {Maravin}, and
  {Kosowsky}}}]{faradaypaper}
\bibinfo{author}{\bibfnamefont{T.}~\bibnamefont{{Kahniashvili}}},
  \bibinfo{author}{\bibfnamefont{Y.}~\bibnamefont{{Maravin}}},
  \bibnamefont{and}
  \bibinfo{author}{\bibfnamefont{A.}~\bibnamefont{{Kosowsky}}},
  \bibinfo{journal}{\prd} \textbf{\bibinfo{volume}{80}}, \bibinfo{eid}{023009}
  (\bibinfo{year}{2009}), \eprint{0806.1876}.

\bibitem[{\citenamefont{{Giovannini} and {Kunze}}(2008)}]{Giovannini:2008}
\bibinfo{author}{\bibfnamefont{M.}~\bibnamefont{{Giovannini}}}
  \bibnamefont{and} \bibinfo{author}{\bibfnamefont{K.~E.}
  \bibnamefont{{Kunze}}}, \bibinfo{journal}{\prd}
  \textbf{\bibinfo{volume}{78}}, \bibinfo{eid}{023010} (\bibinfo{year}{2008}),
  \eprint{0804.3380}.

\bibitem[{\citenamefont{{Finelli} et~al.}(2008)\citenamefont{{Finelli}, {Paci},
  and {Paoletti}}}]{Finelli:2008}
\bibinfo{author}{\bibfnamefont{F.}~\bibnamefont{{Finelli}}},
  \bibinfo{author}{\bibfnamefont{F.}~\bibnamefont{{Paci}}}, \bibnamefont{and}
  \bibinfo{author}{\bibfnamefont{D.}~\bibnamefont{{Paoletti}}},
  \bibinfo{journal}{\prd} \textbf{\bibinfo{volume}{78}},
  \bibinfo{pages}{023510} (\bibinfo{year}{2008}), \eprint{0803.1246}.

\bibitem[{\citenamefont{{Paoletti} et~al.}(2008)\citenamefont{{Paoletti},
  {Finelli}, and {Paci}}}]{Paoletti:2008}
\bibinfo{author}{\bibfnamefont{D.}~\bibnamefont{{Paoletti}}},
  \bibinfo{author}{\bibfnamefont{F.}~\bibnamefont{{Finelli}}},
  \bibnamefont{and} \bibinfo{author}{\bibfnamefont{F.}~\bibnamefont{{Paci}}},
  \bibinfo{journal}{ArXiv e-prints}  (\bibinfo{year}{2008}),
  \eprint{0811.0230}.

\bibitem[{\citenamefont{{Yamazaki}
  et~al.}(2008{\natexlab{a}})\citenamefont{{Yamazaki}, {Ichiki}, {Kajino}, and
  {Mathews}}}]{Yamazaki:2008}
\bibinfo{author}{\bibfnamefont{D.~G.} \bibnamefont{{Yamazaki}}},
  \bibinfo{author}{\bibfnamefont{K.}~\bibnamefont{{Ichiki}}},
  \bibinfo{author}{\bibfnamefont{T.}~\bibnamefont{{Kajino}}}, \bibnamefont{and}
  \bibinfo{author}{\bibfnamefont{G.~J.} \bibnamefont{{Mathews}}},
  \bibinfo{journal}{\prd} \textbf{\bibinfo{volume}{77}},
  \bibinfo{pages}{043005} (\bibinfo{year}{2008}{\natexlab{a}}),
  \eprint{0801.2572}.

\bibitem[{\citenamefont{{Shaw} and {Lewis}}(2010)}]{Shaw:2009}
\bibinfo{author}{\bibfnamefont{J.~R.} \bibnamefont{{Shaw}}} \bibnamefont{and}
  \bibinfo{author}{\bibfnamefont{A.}~\bibnamefont{{Lewis}}},
  \bibinfo{journal}{\prd} \textbf{\bibinfo{volume}{81}},
  \bibinfo{pages}{043517} (\bibinfo{year}{2010}), \eprint{0911.2714}.

\bibitem[{\citenamefont{{Yamazaki}
  et~al.}(2006{\natexlab{a}})\citenamefont{{Yamazaki}, {Ichiki}, {Umezu}, and
  {Hanayama}}}]{Yamazaki:2006b}
\bibinfo{author}{\bibfnamefont{D.~G.} \bibnamefont{{Yamazaki}}},
  \bibinfo{author}{\bibfnamefont{K.}~\bibnamefont{{Ichiki}}},
  \bibinfo{author}{\bibfnamefont{K.}~\bibnamefont{{Umezu}}}, \bibnamefont{and}
  \bibinfo{author}{\bibfnamefont{H.}~\bibnamefont{{Hanayama}}},
  \bibinfo{journal}{\prd} \textbf{\bibinfo{volume}{74}},
  \bibinfo{pages}{123518} (\bibinfo{year}{2006}{\natexlab{a}}),
  \eprint{arXiv:astro-ph/0611910}.

\bibitem[{\citenamefont{{Sethi} et~al.}(2008)\citenamefont{{Sethi}, {Nath}, and
  {Subramanian}}}]{Sethi:2008}
\bibinfo{author}{\bibfnamefont{S.~K.} \bibnamefont{{Sethi}}},
  \bibinfo{author}{\bibfnamefont{B.~B.} \bibnamefont{{Nath}}},
  \bibnamefont{and}
  \bibinfo{author}{\bibfnamefont{K.}~\bibnamefont{{Subramanian}}},
  \bibinfo{journal}{\mnras} \textbf{\bibinfo{volume}{387}},
  \bibinfo{pages}{1589} (\bibinfo{year}{2008}), \eprint{0804.3473}.

\bibitem[{\citenamefont{{Yamazaki}
  et~al.}(2008{\natexlab{b}})\citenamefont{{Yamazaki}, {Ichiki}, {Kajino}, and
  {Mathews}}}]{Yamazaki:2008b}
\bibinfo{author}{\bibfnamefont{D.~G.} \bibnamefont{{Yamazaki}}},
  \bibinfo{author}{\bibfnamefont{K.}~\bibnamefont{{Ichiki}}},
  \bibinfo{author}{\bibfnamefont{T.}~\bibnamefont{{Kajino}}}, \bibnamefont{and}
  \bibinfo{author}{\bibfnamefont{G.~J.} \bibnamefont{{Mathews}}},
  \bibinfo{journal}{\prd} \textbf{\bibinfo{volume}{78}}, \bibinfo{eid}{123001}
  (\bibinfo{year}{2008}{\natexlab{b}}), \eprint{0811.2221}.

\bibitem[{\citenamefont{{Tashiro} and {Sugiyama}}(2011)}]{Tashiro:2009}
\bibinfo{author}{\bibfnamefont{H.}~\bibnamefont{{Tashiro}}} \bibnamefont{and}
  \bibinfo{author}{\bibfnamefont{N.}~\bibnamefont{{Sugiyama}}},
  \bibinfo{journal}{\mnras} \textbf{\bibinfo{volume}{411}},
  \bibinfo{pages}{1284} (\bibinfo{year}{2011}), \eprint{0908.0113}.

\bibitem[{\citenamefont{{Yamazaki}
  et~al.}(2010{\natexlab{a}})\citenamefont{{Yamazaki}, {Ichiki}, {Kajino}, and
  {Mathews}}}]{Yamazaki:2010b}
\bibinfo{author}{\bibfnamefont{D.~G.} \bibnamefont{{Yamazaki}}},
  \bibinfo{author}{\bibfnamefont{K.}~\bibnamefont{{Ichiki}}},
  \bibinfo{author}{\bibfnamefont{T.}~\bibnamefont{{Kajino}}}, \bibnamefont{and}
  \bibinfo{author}{\bibfnamefont{G.~J.} \bibnamefont{{Mathews}}},
  \bibinfo{journal}{\prd} \textbf{\bibinfo{volume}{81}}, \bibinfo{eid}{103519}
  (\bibinfo{year}{2010}{\natexlab{a}}), \eprint{1005.1638}.

\bibitem[{\citenamefont{{Kunze}}(2011)}]{Kunze:2011}
\bibinfo{author}{\bibfnamefont{K.~E.} \bibnamefont{{Kunze}}},
  \bibinfo{journal}{\prd} \textbf{\bibinfo{volume}{83}}, \bibinfo{eid}{023006}
  (\bibinfo{year}{2011}), \eprint{1007.3163}.

\bibitem[{\citenamefont{{Matese} and {O'Connell}}(1970)}]{Matese:1970}
\bibinfo{author}{\bibfnamefont{J.~J.} \bibnamefont{{Matese}}} \bibnamefont{and}
  \bibinfo{author}{\bibfnamefont{R.~F.} \bibnamefont{{O'Connell}}},
  \bibinfo{journal}{\apj} \textbf{\bibinfo{volume}{160}}, \bibinfo{pages}{451}
  (\bibinfo{year}{1970}).

\bibitem[{\citenamefont{{Grasso} and {Rubinstein}}(1996)}]{Grasso:1996}
\bibinfo{author}{\bibfnamefont{D.}~\bibnamefont{{Grasso}}} \bibnamefont{and}
  \bibinfo{author}{\bibfnamefont{H.~R.} \bibnamefont{{Rubinstein}}},
  \bibinfo{journal}{Physics Letters B} \textbf{\bibinfo{volume}{379}},
  \bibinfo{pages}{73} (\bibinfo{year}{1996}), \eprint{arXiv:astro-ph/9602055}.

\bibitem[{\citenamefont{{Caprini} and {Durrer}}(2002)}]{Caprini:2002}
\bibinfo{author}{\bibfnamefont{C.}~\bibnamefont{{Caprini}}} \bibnamefont{and}
  \bibinfo{author}{\bibfnamefont{R.}~\bibnamefont{{Durrer}}},
  \bibinfo{journal}{\prd} \textbf{\bibinfo{volume}{65}},
  \bibinfo{pages}{023517} (\bibinfo{year}{2002}),
  \eprint{arXiv:astro-ph/0106244}.

\bibitem[{\citenamefont{{Dolag} et~al.}(1999)\citenamefont{{Dolag},
  {Bartelmann}, and {Lesch}}}]{Dolag:1999}
\bibinfo{author}{\bibfnamefont{K.}~\bibnamefont{{Dolag}}},
  \bibinfo{author}{\bibfnamefont{M.}~\bibnamefont{{Bartelmann}}},
  \bibnamefont{and} \bibinfo{author}{\bibfnamefont{H.}~\bibnamefont{{Lesch}}},
  \bibinfo{journal}{\aap} \textbf{\bibinfo{volume}{348}}, \bibinfo{pages}{351}
  (\bibinfo{year}{1999}), \eprint{arXiv:astro-ph/0202272}.

\bibitem[{\citenamefont{{Banerjee} and {Jedamzik}}(2004)}]{Banerjee:2004}
\bibinfo{author}{\bibfnamefont{R.}~\bibnamefont{{Banerjee}}} \bibnamefont{and}
  \bibinfo{author}{\bibfnamefont{K.}~\bibnamefont{{Jedamzik}}},
  \bibinfo{journal}{\prd} \textbf{\bibinfo{volume}{70}}, \bibinfo{eid}{123003}
  (\bibinfo{year}{2004}), \eprint{arXiv:astro-ph/0410032}.

\bibitem[{\citenamefont{{Yamazaki}
  et~al.}(2006{\natexlab{b}})\citenamefont{{Yamazaki}, {Ichiki}, {Kajino}, and
  {Mathews}}}]{Yamazaki:2006}
\bibinfo{author}{\bibfnamefont{D.~G.} \bibnamefont{{Yamazaki}}},
  \bibinfo{author}{\bibfnamefont{K.}~\bibnamefont{{Ichiki}}},
  \bibinfo{author}{\bibfnamefont{T.}~\bibnamefont{{Kajino}}}, \bibnamefont{and}
  \bibinfo{author}{\bibfnamefont{G.~J.} \bibnamefont{{Mathews}}},
  \bibinfo{journal}{\apj} \textbf{\bibinfo{volume}{646}}, \bibinfo{pages}{719}
  (\bibinfo{year}{2006}{\natexlab{b}}), \eprint{arXiv:astro-ph/0602224}.

\bibitem[{\citenamefont{{Yamazaki}
  et~al.}(2010{\natexlab{b}})\citenamefont{{Yamazaki}, {Ichiki}, {Kajino}, and
  {Mathews}}}]{Yamazaki:2010}
\bibinfo{author}{\bibfnamefont{D.~G.} \bibnamefont{{Yamazaki}}},
  \bibinfo{author}{\bibfnamefont{K.}~\bibnamefont{{Ichiki}}},
  \bibinfo{author}{\bibfnamefont{T.}~\bibnamefont{{Kajino}}}, \bibnamefont{and}
  \bibinfo{author}{\bibfnamefont{G.~J.} \bibnamefont{{Mathews}}},
  \bibinfo{journal}{\prd} \textbf{\bibinfo{volume}{81}},
  \bibinfo{pages}{023008} (\bibinfo{year}{2010}{\natexlab{b}}),
  \eprint{1001.2012}.

\bibitem[{\citenamefont{{Paoletti} and {Finelli}}(2010)}]{Paoletti:2010}
\bibinfo{author}{\bibfnamefont{D.}~\bibnamefont{{Paoletti}}} \bibnamefont{and}
  \bibinfo{author}{\bibfnamefont{F.}~\bibnamefont{{Finelli}}},
  \bibinfo{journal}{ArXiv e-prints}  (\bibinfo{year}{2010}),
  \eprint{1005.0148}.

\bibitem[{\citenamefont{{Brown} and {Crittenden}}(2005)}]{Brown:2005}
\bibinfo{author}{\bibfnamefont{I.}~\bibnamefont{{Brown}}} \bibnamefont{and}
  \bibinfo{author}{\bibfnamefont{R.}~\bibnamefont{{Crittenden}}},
  \bibinfo{journal}{\prd} \textbf{\bibinfo{volume}{72}},
  \bibinfo{pages}{063002} (\bibinfo{year}{2005}),
  \eprint{arXiv:astro-ph/0506570}.

\bibitem[{\citenamefont{{Seshadri} and {Subramanian}}(2009)}]{Seshadri:2009}
\bibinfo{author}{\bibfnamefont{T.~R.} \bibnamefont{{Seshadri}}}
  \bibnamefont{and}
  \bibinfo{author}{\bibfnamefont{K.}~\bibnamefont{{Subramanian}}},
  \bibinfo{journal}{Physical Review Letters} \textbf{\bibinfo{volume}{103}},
  \bibinfo{pages}{081303} (\bibinfo{year}{2009}), \eprint{0902.4066}.

\bibitem[{\citenamefont{{Caprini} et~al.}(2009)\citenamefont{{Caprini},
  {Finelli}, {Paoletti}, and {Riotto}}}]{caprini:2009b}
\bibinfo{author}{\bibfnamefont{C.}~\bibnamefont{{Caprini}}},
  \bibinfo{author}{\bibfnamefont{F.}~\bibnamefont{{Finelli}}},
  \bibinfo{author}{\bibfnamefont{D.}~\bibnamefont{{Paoletti}}},
  \bibnamefont{and} \bibinfo{author}{\bibfnamefont{A.}~\bibnamefont{{Riotto}}},
  \bibinfo{journal}{\jcap} \textbf{\bibinfo{volume}{6}}, \bibinfo{pages}{21}
  (\bibinfo{year}{2009}), \eprint{0903.1420}.

\bibitem[{\citenamefont{{Durrer}}(1994)}]{Durrer:1994}
\bibinfo{author}{\bibfnamefont{R.}~\bibnamefont{{Durrer}}},
  \bibinfo{journal}{Fundamentals of Cosmic Physics}
  \textbf{\bibinfo{volume}{15}}, \bibinfo{pages}{209} (\bibinfo{year}{1994}),
  \eprint{arXiv:astro-ph/9311041}.

\bibitem[{\citenamefont{{Durrer} and {Straumann}}(1988)}]{Durrer:1988}
\bibinfo{author}{\bibfnamefont{R.}~\bibnamefont{{Durrer}}} \bibnamefont{and}
  \bibinfo{author}{\bibfnamefont{N.}~\bibnamefont{{Straumann}}},
  \bibinfo{journal}{Helv. Phys. Acta} \textbf{\bibinfo{volume}{61}},
  \bibinfo{pages}{1027} (\bibinfo{year}{1988}),
  \urlprefix\url{http://theory.physics.unige.ch/~durrer/papers/threeplusone.pdf}.

\bibitem[{\citenamefont{{Subramanian} and
  {Barrow}}(1998{\natexlab{b}})}]{Subramanian:1998}
\bibinfo{author}{\bibfnamefont{K.}~\bibnamefont{{Subramanian}}}
  \bibnamefont{and} \bibinfo{author}{\bibfnamefont{J.~D.}
  \bibnamefont{{Barrow}}}, \bibinfo{journal}{\prd}
  \textbf{\bibinfo{volume}{58}}, \bibinfo{pages}{083502}
  (\bibinfo{year}{1998}{\natexlab{b}}), \eprint{arXiv:astro-ph/9712083}.

\bibitem[{\citenamefont{{Jedamzik} et~al.}(1998)\citenamefont{{Jedamzik},
  {Katalini{\'c}}, and {Olinto}}}]{Jedamzik:1998}
\bibinfo{author}{\bibfnamefont{K.}~\bibnamefont{{Jedamzik}}},
  \bibinfo{author}{\bibfnamefont{V.}~\bibnamefont{{Katalini{\'c}}}},
  \bibnamefont{and} \bibinfo{author}{\bibfnamefont{A.~V.}
  \bibnamefont{{Olinto}}}, \bibinfo{journal}{\prd}
  \textbf{\bibinfo{volume}{57}}, \bibinfo{pages}{3264} (\bibinfo{year}{1998}),
  \eprint{arXiv:astro-ph/9606080}.

\bibitem[{\citenamefont{{Kim} et~al.}(1996)\citenamefont{{Kim}, {Olinto}, and
  {Rosner}}}]{Kim:1996}
\bibinfo{author}{\bibfnamefont{E.}~\bibnamefont{{Kim}}},
  \bibinfo{author}{\bibfnamefont{A.~V.} \bibnamefont{{Olinto}}},
  \bibnamefont{and} \bibinfo{author}{\bibfnamefont{R.}~\bibnamefont{{Rosner}}},
  \bibinfo{journal}{\apj} \textbf{\bibinfo{volume}{468}}, \bibinfo{pages}{28}
  (\bibinfo{year}{1996}), \eprint{arXiv:astro-ph/9412070}.

\bibitem[{\citenamefont{{Barrow} et~al.}(2007)\citenamefont{{Barrow},
  {Maartens}, and {Tsagas}}}]{Barrow:2007}
\bibinfo{author}{\bibfnamefont{J.~D.} \bibnamefont{{Barrow}}},
  \bibinfo{author}{\bibfnamefont{R.}~\bibnamefont{{Maartens}}},
  \bibnamefont{and} \bibinfo{author}{\bibfnamefont{C.~G.}
  \bibnamefont{{Tsagas}}}, \bibinfo{journal}{\physrep}
  \textbf{\bibinfo{volume}{449}}, \bibinfo{pages}{131} (\bibinfo{year}{2007}),
  \eprint{arXiv:astro-ph/0611537}.

\bibitem[{\citenamefont{{Adams} et~al.}(1996)\citenamefont{{Adams},
  {Danielsson}, {Grasso}, and {Rubinstein}}}]{Adams:1996}
\bibinfo{author}{\bibfnamefont{J.}~\bibnamefont{{Adams}}},
  \bibinfo{author}{\bibfnamefont{U.~H.} \bibnamefont{{Danielsson}}},
  \bibinfo{author}{\bibfnamefont{D.}~\bibnamefont{{Grasso}}}, \bibnamefont{and}
  \bibinfo{author}{\bibfnamefont{H.}~\bibnamefont{{Rubinstein}}},
  \bibinfo{journal}{Physics Letters B} \textbf{\bibinfo{volume}{388}},
  \bibinfo{pages}{253} (\bibinfo{year}{1996}), \eprint{arXiv:astro-ph/9607043}.

\bibitem[{\citenamefont{{Sethi} and {Subramanian}}(2005)}]{Sethi:2004}
\bibinfo{author}{\bibfnamefont{S.~K.} \bibnamefont{{Sethi}}} \bibnamefont{and}
  \bibinfo{author}{\bibfnamefont{K.}~\bibnamefont{{Subramanian}}},
  \bibinfo{journal}{\mnras} \textbf{\bibinfo{volume}{356}},
  \bibinfo{pages}{778} (\bibinfo{year}{2005}), \eprint{arXiv:astro-ph/0405413}.

\bibitem[{\citenamefont{{Lewis} et~al.}(2000)\citenamefont{{Lewis},
  {Challinor}, and {Lasenby}}}]{CAMB}
\bibinfo{author}{\bibfnamefont{A.}~\bibnamefont{{Lewis}}},
  \bibinfo{author}{\bibfnamefont{A.}~\bibnamefont{{Challinor}}},
  \bibnamefont{and}
  \bibinfo{author}{\bibfnamefont{A.}~\bibnamefont{{Lasenby}}},
  \bibinfo{journal}{\apj} \textbf{\bibinfo{volume}{538}}, \bibinfo{pages}{473}
  (\bibinfo{year}{2000}), \eprint{arXiv:astro-ph/9911177}.

\bibitem[{\citenamefont{Lewis and Bridle}(2002)}]{COSMOMC}
\bibinfo{author}{\bibfnamefont{A.}~\bibnamefont{Lewis}} \bibnamefont{and}
  \bibinfo{author}{\bibfnamefont{S.}~\bibnamefont{Bridle}},
  \bibinfo{journal}{Phys. Rev.} \textbf{\bibinfo{volume}{D66}},
  \bibinfo{pages}{103511} (\bibinfo{year}{2002}), \eprint{astro-ph/0205436}.

\bibitem[{\citenamefont{{Jarosik} et~al.}(2010)\citenamefont{{Jarosik},
  {Bennett}, {Dunkley}, {Gold}, {Greason}, {Halpern}, {Hill}, {Hinshaw},
  {Kogut}, {Komatsu} et~al.}}]{WMAP7}
\bibinfo{author}{\bibfnamefont{N.}~\bibnamefont{{Jarosik}}},
  \bibinfo{author}{\bibfnamefont{C.~L.} \bibnamefont{{Bennett}}},
  \bibinfo{author}{\bibfnamefont{J.}~\bibnamefont{{Dunkley}}},
  \bibinfo{author}{\bibfnamefont{B.}~\bibnamefont{{Gold}}},
  \bibinfo{author}{\bibfnamefont{M.~R.} \bibnamefont{{Greason}}},
  \bibinfo{author}{\bibfnamefont{M.}~\bibnamefont{{Halpern}}},
  \bibinfo{author}{\bibfnamefont{R.~S.} \bibnamefont{{Hill}}},
  \bibinfo{author}{\bibfnamefont{G.}~\bibnamefont{{Hinshaw}}},
  \bibinfo{author}{\bibfnamefont{A.}~\bibnamefont{{Kogut}}},
  \bibinfo{author}{\bibfnamefont{E.}~\bibnamefont{{Komatsu}}},
  \bibnamefont{et~al.}, \bibinfo{journal}{ArXiv e-prints}
  (\bibinfo{year}{2010}), \eprint{1001.4744}.

\bibitem[{\citenamefont{{Brown} et~al.}(2009)\citenamefont{{Brown}, {Ade},
  {Bock}, {Bowden}, {Cahill}, {Castro}, {Church}, {Culverhouse}, {Friedman},
  {Ganga} et~al.}}]{QUAD}
\bibinfo{author}{\bibfnamefont{M.~L.} \bibnamefont{{Brown}}},
  \bibinfo{author}{\bibfnamefont{P.}~\bibnamefont{{Ade}}},
  \bibinfo{author}{\bibfnamefont{J.}~\bibnamefont{{Bock}}},
  \bibinfo{author}{\bibfnamefont{M.}~\bibnamefont{{Bowden}}},
  \bibinfo{author}{\bibfnamefont{G.}~\bibnamefont{{Cahill}}},
  \bibinfo{author}{\bibfnamefont{P.~G.} \bibnamefont{{Castro}}},
  \bibinfo{author}{\bibfnamefont{S.}~\bibnamefont{{Church}}},
  \bibinfo{author}{\bibfnamefont{T.}~\bibnamefont{{Culverhouse}}},
  \bibinfo{author}{\bibfnamefont{R.~B.} \bibnamefont{{Friedman}}},
  \bibinfo{author}{\bibfnamefont{K.}~\bibnamefont{{Ganga}}},
  \bibnamefont{et~al.}, \bibinfo{journal}{\apj} \textbf{\bibinfo{volume}{705}},
  \bibinfo{pages}{978} (\bibinfo{year}{2009}), \eprint{0906.1003}.

\bibitem[{\citenamefont{{Reichardt} et~al.}(2009)\citenamefont{{Reichardt},
  {Ade}, {Bock}, {Bond}, {Brevik}, {Contaldi}, {Daub}, {Dempsey}, {Goldstein},
  {Holzapfel} et~al.}}]{ACBAR}
\bibinfo{author}{\bibfnamefont{C.~L.} \bibnamefont{{Reichardt}}},
  \bibinfo{author}{\bibfnamefont{P.~A.~R.} \bibnamefont{{Ade}}},
  \bibinfo{author}{\bibfnamefont{J.~J.} \bibnamefont{{Bock}}},
  \bibinfo{author}{\bibfnamefont{J.~R.} \bibnamefont{{Bond}}},
  \bibinfo{author}{\bibfnamefont{J.~A.} \bibnamefont{{Brevik}}},
  \bibinfo{author}{\bibfnamefont{C.~R.} \bibnamefont{{Contaldi}}},
  \bibinfo{author}{\bibfnamefont{M.~D.} \bibnamefont{{Daub}}},
  \bibinfo{author}{\bibfnamefont{J.~T.} \bibnamefont{{Dempsey}}},
  \bibinfo{author}{\bibfnamefont{J.~H.} \bibnamefont{{Goldstein}}},
  \bibinfo{author}{\bibfnamefont{W.~L.} \bibnamefont{{Holzapfel}}},
  \bibnamefont{et~al.}, \bibinfo{journal}{\apj} \textbf{\bibinfo{volume}{694}},
  \bibinfo{pages}{1200} (\bibinfo{year}{2009}), \eprint{0801.1491}.

\bibitem[{\citenamefont{Freedman et~al.}(2001)}]{HubbleKey}
\bibinfo{author}{\bibfnamefont{W.~L.} \bibnamefont{Freedman}}
  \bibnamefont{et~al.}, \bibinfo{journal}{Astrophys. J.}
  \textbf{\bibinfo{volume}{553}}, \bibinfo{pages}{47} (\bibinfo{year}{2001}),
  \eprint{astro-ph/0012376}.

\bibitem[{\citenamefont{Kowalski et~al.}(2008)}]{SupernovaUnion}
\bibinfo{author}{\bibfnamefont{M.}~\bibnamefont{Kowalski}} \bibnamefont{et~al.}
  (\bibinfo{year}{2008}), \eprint{0804.4142}.

\bibitem[{\citenamefont{Percival et~al.}(2009)}]{BAO}
\bibinfo{author}{\bibfnamefont{W.~J.} \bibnamefont{Percival}}
  \bibnamefont{et~al.} (\bibinfo{year}{2009}), \eprint{0907.1660}.

\bibitem[{\citenamefont{Cole et~al.}(2005)}]{2dF}
\bibinfo{author}{\bibfnamefont{S.}~\bibnamefont{Cole}} \bibnamefont{et~al.}
  (\bibinfo{collaboration}{The 2dFGRS}), \bibinfo{journal}{Mon. Not. Roy.
  Astron. Soc.} \textbf{\bibinfo{volume}{362}}, \bibinfo{pages}{505}
  (\bibinfo{year}{2005}), \eprint{astro-ph/0501174}.

\bibitem[{\citenamefont{Tegmark et~al.}(2006)}]{SDSS4}
\bibinfo{author}{\bibfnamefont{M.}~\bibnamefont{Tegmark}} \bibnamefont{et~al.},
  \bibinfo{journal}{Phys. Rev.} \textbf{\bibinfo{volume}{D74}},
  \bibinfo{pages}{123507} (\bibinfo{year}{2006}), \eprint{astro-ph/0608632}.

\bibitem[{\citenamefont{{Reid} et~al.}(2010)\citenamefont{{Reid}, {Percival},
  {Eisenstein}, {Verde}, {Spergel}, {Skibba}, {Bahcall}, {Budavari}, {Frieman},
  {Fukugita} et~al.}}]{SDSS7}
\bibinfo{author}{\bibfnamefont{B.~A.} \bibnamefont{{Reid}}},
  \bibinfo{author}{\bibfnamefont{W.~J.} \bibnamefont{{Percival}}},
  \bibinfo{author}{\bibfnamefont{D.~J.} \bibnamefont{{Eisenstein}}},
  \bibinfo{author}{\bibfnamefont{L.}~\bibnamefont{{Verde}}},
  \bibinfo{author}{\bibfnamefont{D.~N.} \bibnamefont{{Spergel}}},
  \bibinfo{author}{\bibfnamefont{R.~A.} \bibnamefont{{Skibba}}},
  \bibinfo{author}{\bibfnamefont{N.~A.} \bibnamefont{{Bahcall}}},
  \bibinfo{author}{\bibfnamefont{T.}~\bibnamefont{{Budavari}}},
  \bibinfo{author}{\bibfnamefont{J.~A.} \bibnamefont{{Frieman}}},
  \bibinfo{author}{\bibfnamefont{M.}~\bibnamefont{{Fukugita}}},
  \bibnamefont{et~al.}, \bibinfo{journal}{\mnras}
  \textbf{\bibinfo{volume}{404}}, \bibinfo{pages}{60} (\bibinfo{year}{2010}),
  \eprint{0907.1659}.

\bibitem[{\citenamefont{{Lueker} et~al.}(2009)\citenamefont{{Lueker},
  {Reichardt}, {Schaffer}, {Zahn}, {Ade}, {Aird}, {Benson}, {Bleem},
  {Carlstrom}, {Chang} et~al.}}]{SPT}
\bibinfo{author}{\bibfnamefont{M.}~\bibnamefont{{Lueker}}},
  \bibinfo{author}{\bibfnamefont{C.~L.} \bibnamefont{{Reichardt}}},
  \bibinfo{author}{\bibfnamefont{K.~K.} \bibnamefont{{Schaffer}}},
  \bibinfo{author}{\bibfnamefont{O.}~\bibnamefont{{Zahn}}},
  \bibinfo{author}{\bibfnamefont{P.~A.~R.} \bibnamefont{{Ade}}},
  \bibinfo{author}{\bibfnamefont{K.~A.} \bibnamefont{{Aird}}},
  \bibinfo{author}{\bibfnamefont{B.~A.} \bibnamefont{{Benson}}},
  \bibinfo{author}{\bibfnamefont{L.~E.} \bibnamefont{{Bleem}}},
  \bibinfo{author}{\bibfnamefont{J.~E.} \bibnamefont{{Carlstrom}}},
  \bibinfo{author}{\bibfnamefont{C.~L.} \bibnamefont{{Chang}}},
  \bibnamefont{et~al.}, \bibinfo{journal}{ArXiv e-prints}
  (\bibinfo{year}{2009}), \eprint{0912.4317}.

\bibitem[{\citenamefont{{Komatsu} and {Seljak}}(2002)}]{Komatsu:2002}
\bibinfo{author}{\bibfnamefont{E.}~\bibnamefont{{Komatsu}}} \bibnamefont{and}
  \bibinfo{author}{\bibfnamefont{U.}~\bibnamefont{{Seljak}}},
  \bibinfo{journal}{\mnras} \textbf{\bibinfo{volume}{336}},
  \bibinfo{pages}{1256} (\bibinfo{year}{2002}),
  \eprint{arXiv:astro-ph/0205468}.

\bibitem[{\citenamefont{Zhang and Sheth}(2007)}]{Zhang:2007}
\bibinfo{author}{\bibfnamefont{P.}~\bibnamefont{Zhang}} \bibnamefont{and}
  \bibinfo{author}{\bibfnamefont{R.~K.} \bibnamefont{Sheth}},
  \bibinfo{journal}{Astrophys. J.} \textbf{\bibinfo{volume}{671}},
  \bibinfo{pages}{14} (\bibinfo{year}{2007}), \eprint{astro-ph/0701879}.

\bibitem[{\citenamefont{{McDonald} et~al.}(2005)\citenamefont{{McDonald},
  {Seljak}, {Cen}, {Shih}, {Weinberg}, {Burles}, {Schneider}, {Schlegel},
  {Bahcall}, {Briggs} et~al.}}]{McDonald:2005}
\bibinfo{author}{\bibfnamefont{P.}~\bibnamefont{{McDonald}}},
  \bibinfo{author}{\bibfnamefont{U.}~\bibnamefont{{Seljak}}},
  \bibinfo{author}{\bibfnamefont{R.}~\bibnamefont{{Cen}}},
  \bibinfo{author}{\bibfnamefont{D.}~\bibnamefont{{Shih}}},
  \bibinfo{author}{\bibfnamefont{D.~H.} \bibnamefont{{Weinberg}}},
  \bibinfo{author}{\bibfnamefont{S.}~\bibnamefont{{Burles}}},
  \bibinfo{author}{\bibfnamefont{D.~P.} \bibnamefont{{Schneider}}},
  \bibinfo{author}{\bibfnamefont{D.~J.} \bibnamefont{{Schlegel}}},
  \bibinfo{author}{\bibfnamefont{N.~A.} \bibnamefont{{Bahcall}}},
  \bibinfo{author}{\bibfnamefont{J.~W.} \bibnamefont{{Briggs}}},
  \bibnamefont{et~al.}, \bibinfo{journal}{\apj} \textbf{\bibinfo{volume}{635}},
  \bibinfo{pages}{761} (\bibinfo{year}{2005}), \eprint{arXiv:astro-ph/0407377}.

\bibitem[{\citenamefont{{Maggiore}}(2000)}]{Maggiore:2000}
\bibinfo{author}{\bibfnamefont{M.}~\bibnamefont{{Maggiore}}},
  \bibinfo{journal}{ArXiv General Relativity and Quantum Cosmology e-prints}
  (\bibinfo{year}{2000}), \eprint{arXiv:gr-qc/0008027}.

\bibitem[{\citenamefont{{Caprini} and {Durrer}}(2005)}]{Caprini:2005}
\bibinfo{author}{\bibfnamefont{C.}~\bibnamefont{{Caprini}}} \bibnamefont{and}
  \bibinfo{author}{\bibfnamefont{R.}~\bibnamefont{{Durrer}}},
  \bibinfo{journal}{\prd} \textbf{\bibinfo{volume}{72}}, \bibinfo{eid}{088301}
  (\bibinfo{year}{2005}), \eprint{arXiv:astro-ph/0504553}.

\bibitem[{\citenamefont{{Navarro} et~al.}(1997)\citenamefont{{Navarro}, Frenk,
  and White}}]{NFW}
\bibinfo{author}{\bibfnamefont{J.~F.} \bibnamefont{{Navarro}}},
  \bibinfo{author}{\bibfnamefont{C.~S.} \bibnamefont{Frenk}}, \bibnamefont{and}
  \bibinfo{author}{\bibfnamefont{S.~D.~M.} \bibnamefont{White}},
  \bibinfo{journal}{Astrophys. J.} \textbf{\bibinfo{volume}{490}},
  \bibinfo{pages}{493} (\bibinfo{year}{1997}), \eprint{astro-ph/9611107}.

\bibitem[{\citenamefont{{Bryan} and {Norman}}(1998)}]{Bryan:1998}
\bibinfo{author}{\bibfnamefont{G.~L.} \bibnamefont{{Bryan}}} \bibnamefont{and}
  \bibinfo{author}{\bibfnamefont{M.~L.} \bibnamefont{{Norman}}},
  \bibinfo{journal}{\apj} \textbf{\bibinfo{volume}{495}}, \bibinfo{pages}{80}
  (\bibinfo{year}{1998}), \eprint{arXiv:astro-ph/9710107}.

\bibitem[{\citenamefont{{Duffy} et~al.}(2008)\citenamefont{{Duffy}, {Schaye},
  {Kay}, and {Dalla Vecchia}}}]{Duffy:2008}
\bibinfo{author}{\bibfnamefont{A.~R.} \bibnamefont{{Duffy}}},
  \bibinfo{author}{\bibfnamefont{J.}~\bibnamefont{{Schaye}}},
  \bibinfo{author}{\bibfnamefont{S.~T.} \bibnamefont{{Kay}}}, \bibnamefont{and}
  \bibinfo{author}{\bibfnamefont{C.}~\bibnamefont{{Dalla Vecchia}}},
  \bibinfo{journal}{\mnras} \textbf{\bibinfo{volume}{390}},
  \bibinfo{pages}{L64} (\bibinfo{year}{2008}), \eprint{0804.2486}.

\bibitem[{\citenamefont{{Jenkins} et~al.}(2001)\citenamefont{{Jenkins},
  {Frenk}, {White}, {Colberg}, {Cole}, {Evrard}, {Couchman}, and
  {Yoshida}}}]{Jenkins}
\bibinfo{author}{\bibfnamefont{A.}~\bibnamefont{{Jenkins}}},
  \bibinfo{author}{\bibfnamefont{C.~S.} \bibnamefont{{Frenk}}},
  \bibinfo{author}{\bibfnamefont{S.~D.~M.} \bibnamefont{{White}}},
  \bibinfo{author}{\bibfnamefont{J.~M.} \bibnamefont{{Colberg}}},
  \bibinfo{author}{\bibfnamefont{S.}~\bibnamefont{{Cole}}},
  \bibinfo{author}{\bibfnamefont{A.~E.} \bibnamefont{{Evrard}}},
  \bibinfo{author}{\bibfnamefont{H.~M.~P.} \bibnamefont{{Couchman}}},
  \bibnamefont{and}
  \bibinfo{author}{\bibfnamefont{N.}~\bibnamefont{{Yoshida}}},
  \bibinfo{journal}{\mnras} \textbf{\bibinfo{volume}{321}},
  \bibinfo{pages}{372} (\bibinfo{year}{2001}), \eprint{arXiv:astro-ph/0005260}.

\end{thebibliography}

\end{document}